\documentclass{article}

\usepackage{arxiv}

\usepackage[utf8]{inputenc} 
\usepackage[T1]{fontenc}    
\usepackage{hyperref}       
\usepackage{url}            
\usepackage{booktabs}       
\usepackage{amsfonts}       
\usepackage{nicefrac}       
\usepackage{microtype}      
\usepackage{lipsum}
\usepackage{graphicx}

\usepackage{scalerel}
\usepackage{tikz}
\usetikzlibrary{svg.path}
\usepackage{cite}
\usepackage{amsmath,amssymb,amsfonts}
\usepackage{textcomp}
\usepackage{multirow}
\usepackage[linesnumbered, longend, algoruled]{algorithm2e}
\usepackage{soul}

\DeclareMathOperator*{\argmin}{\textbf{argmin}} 
\usepackage{textcomp}
\usepackage{cite}
\usepackage{xcolor}
\usepackage{tabularx}
\usepackage{amsthm}
\usepackage{mathtools}
\usepackage{caption}
\usepackage{subcaption}
\usepackage{algorithmic}
\usepackage{algorithm2e}

\definecolor{orcidlogocol}{HTML}{A6CE39}
\tikzset{
  orcidlogo/.pic={
    \fill[orcidlogocol] svg{M256,128c0,70.7-57.3,128-128,128C57.3,256,0,198.7,0,128C0,57.3,57.3,0,128,0C198.7,0,256,57.3,256,128z};
    \fill[white] svg{M86.3,186.2H70.9V79.1h15.4v48.4V186.2z}
                 svg{M108.9,79.1h41.6c39.6,0,57,28.3,57,53.6c0,27.5-21.5,53.6-56.8,53.6h-41.8V79.1z M124.3,172.4h24.5c34.9,0,42.9-26.5,42.9-39.7c0-21.5-13.7-39.7-43.7-39.7h-23.7V172.4z}
                 svg{M88.7,56.8c0,5.5-4.5,10.1-10.1,10.1c-5.6,0-10.1-4.6-10.1-10.1c0-5.6,4.5-10.1,10.1-10.1C84.2,46.7,88.7,51.3,88.7,56.8z};
  }
}

\newcommand\orcidicon[1]{\href{https://orcid.org/#1}{\mbox{\scalerel*{
\begin{tikzpicture}[yscale=-1,transform shape]
\pic{orcidlogo};
\end{tikzpicture}
}{|}}}}

\usepackage{hyperref} 

\begin{document}

\title{ReCo1: A Fault resilient technique of Correlation Sensitive Stochastic Designs
\thanks{}
}

\author{ Shyamali Mitra\orcidicon{}, Sayantan Banerjee\orcidicon{}, Mrinal Kanti Naskar\orcidicon{},
\thanks{ S. Mitra is with the Dept. of IEE , Jadavpur University, India (e-mail: shyamalimitra.iee@jadavpuruniversity.in ) }
\thanks{ S Banerjee is with the Dept. of IEE , Jadavpur University, India (e-mail: sayantan099@gmail.com )}
\thanks{M. K. Naskar is with the Dept. of ETCE , Jadavpur University, India (e-mail: mrinaletce@gmail.com )}
}

\markboth{Journal of \LaTeX\ Class Files,~Vol.~14, No.~8, August~2015}%
{Shell \MakeLowercase{\textit{et al.}}: Bare Demo of IEEEtran.cls for IEEE Journals}

\maketitle 

\begin{abstract}
In stochastic circuits, major sources of error are correlation errors, soft errors and random fluctuation errors that affect the accuracy and reliability of the circuit. The soft error has the effect of changing the  correlation status and in turn changes the probability of numbers leading to the erroneous output. This has serious impact on security and medical systems where highly accurate systems are required. We tackle this problem by introducing the fault-tolerant technique of correlation-sensitive stochastic logic circuits. We develop a framework of \textit{Remodelling Correlation}  (\textit{ReCo}) for Stochastic Logic Elements; AND, XOR and OR for reliable operation.  We present two variants of \textit{ReCo} models in combinational circuits with contradictory requirements by stating two interesting case studies.  The proposed technique selects logic elements and places correction blocks based on a priority-based rule that helps to converge to the desired MSE quickly requiring less hardware area.  It is shown that this technique does not alter the reliability of the overall circuit. To demonstrate the practical effectiveness of the proposed framework, contrast stretch operation on a standard image in a noisy environment is studied. A  high structural similarity index measure of $92.80$ is observed for the output image with the proposed approach compared to the image (with error) $66.43$.

\end{abstract}

\begin{keywords} {Stochastic Logic Circuits, Soft errors, Reliability, Remodelling Correlation (ReCo) Framework,
 Priority-based approach.}
\end{keywords}

\section{Introduction}
Computation on binary numbers using \textit{Stochastic computing} \cite{gaines1967stochastic} is gaining popularity nowadays because it offers several advantages \cite{alaghi2013survey} compared to conventional weighted-binary computation. It is a low power and low cost alternative to complex arithmetic functions. With a remarkable reduction in size, circuit complexity and power consumption, stochastic architecture has proved to be noise immune compared to the conventional implementation of binarization algorithms \cite{najafi2015fast} and various other image processing tasks also \cite{li2011using}. Though different types of errors such as soft errors, correlation induced errors and random fluctuation errors are identified that affect the accuracy and reliability of stochastic circuits \cite{alaghi2017promise}.  Thus to generate the desired function using unreliable components in presence of these errors has become a challenging task. 
\par  There are several analytical approaches to  assess the reliability of probabilistic circuits with unpredictable behaviours; Probabilistic Gate Models (PGM), Probabilistic Transfer Matrices (PTM), Stochastic computational Model (SCM), Monte Carlo Simulation\cite{xiao2014gate} etc. During analysis we have extensively used PTM throughout the paper. Using PTM, we  showed that though the accuracy of an adder using MUX is controlled by select inputs, they must be taken in order with  inputs to reduce error. Nevertheless to mention that majority of works are aligned towards reliability assessment and analysis \cite{Vallero2019,Xu2019}.
\par  Transient or soft errors are caused due to exposure to external radiation  and are greatly increased by manufacturing defects in a chip. These introduce false logic at the output of the circuit \cite{nicolaidis2005design}. Thus, if multiple faults strike nodes of a gate, the output may be obtained erroneously. Soft errors  are responsible for bit-flips in stochastic bitstream and might induce an undesired correlation between two numbers. This paper emphasizes the inaccurate behaviour of stochastic circuits, contributed mainly due to transient errors under noisy conditions and its methodical correction using correlation in a positive manner.
The present study is dedicated to analyzing the behaviour of  circuits that are subjected to soft errors. Desired correlations are injected into bitstreams to diminish the effect of transient faults and ensure reliable operation of the circuit.  Experiments are conducted on SLEs that are susceptible to changes in correlation. The goal is to create a technology independent framework for complex circuits  to observe error-free output using minimum hardware. Studies are also conducted to show that the reliability of a circuit is not affected by correlation alteration between inputs. The contributions in the present work are highlighted as follows:
\begin{itemize}
  \item We develop a correlation-based framework for correlation sensitive SLEs under transient error scenarios to model the error-free output.  \item A  priority-based approach in the selection of SLEs is explored to reduce the hardware complexity and improve the accuracy in computation for complex circuits.
 \item A study on the effect of the proposed framework on reliability of unreliable circuits.  \item Evaluation of the proposed work on contrast stretch operations on image under  high transient error rates. 
\end{itemize}
In essence, this paper not only contradicts the popular perception with regard to correlation, but firmly establishes that injecting controlled degree of correlation can improve error-resilient behaviour of the circuit. The rest of the paper is organized as follows: In section II, we have analyzed the erroneous behaviour of the multiplexer circuit in the light of PTM. Section III discusses the two major sources of error in stochastic circuits and introduces the proposed methodology in the noisy environment for correlation-sensitive SLEs. With several initial correlation assumptions, we can establish an operating point of the circuit with a suitably injected correlation that could generate an accurate result under the stated conditions. In Section IV, we  extended the idea to simulate complex SLCs to show the effectiveness of the proposed model. Two approaches based on a priority-search model are demonstrated following two distinct conditions. Applicability of the proposed methodology in the context of an image processing task is discussed in Section V. In Section VI, highlights of the experimental results are jotted down with the pros and cons of the proposed algorithms.
\section{USING PTM FOR ANALYSING STOCHASTIC CIRCUITS}
 The PTM, which is used in analysing probabilistic logic circuits\cite{krishnaswamy2008probabilistic} has proved  to be a  convenient tool in error\cite{alaghi2019accuracy} and reliability analysis \cite{krishnaswamy2005accurate} of small stochastic circuits. It can be observed as a conditional probability matrix $M$, so that, $M(i,j)=p(output=j|input=i)$, where $p$ represents the conditional probability of a particular output being true given a certain input combination. For large circuits computation with PTM is tedious. For a circuit with $k$ inputs and $l$ outputs, a circuit PTM is of size $2^k\times2^l$.  In Ideal Transfer Matrix (ITM), when the gate is assumed to be error free, elements are either 0 or 1, representing exact binary values in place of probabilities. PTM  and ITM for a two input AND gate are represented as matrices $J$ and $M$.
\\\begin{equation}\label{eq3}\nonumber\footnotesize
J =
\begin{bmatrix}
1 & 0\\
1 & 0\\
1 & 0\\
0 & 1
\end{bmatrix}
  \hspace{5mm}M =
\begin{bmatrix} \footnotesize
1-p_e & p_e \\
1-p_e & p_e \\
1-p_e & p_e \\
p_e & 1-p_e 
\end{bmatrix}
\end{equation}
The rows in  $J$ and $M$ correspond to input combinations $00,01,10$,$11$. Columns correspond to outputs $0,1$.
The  PTM for large circuits is computed using two basic operations \cite{krishnaswamy2008probabilistic}:
\begin{itemize}
 \item The overall PTM of two or more gates with PTMs $M_1,M_2,...,M_p$  connected in series, is obtained by multiplication of individual PTM; $M_{series} = M_1.M_2...M_p$.
  \item  The resultant PTM of gates with PTMs $M_1,M_2,...,M_p$ connected in parallel is obtained by the Kronecker product of individual PTM; $M_{parallel} = M_1\otimes M_2\otimes M_3...M_p$.
 \end{itemize}
PTM can be defined for a  single gate as well as for the whole circuit, e.g., the accurate condition for addition and subtraction using MUX can be demonstrated using PTM analysis.
\par a) \textit{Deriving condition for accurate Stochastic Addition:} It is reported that the multiplexer performs the addition operation irrespective of the correlation between input bitstreams $A$ and $B$ with the probability of select line being $p_s=0.5$ \cite{alaghi2015functions}. But  the combinations of input bitstreams limit the accuracy. We show that by using PTM  analysis of MUX.
\par Consider a $2:1$ MUX whose output is given as  $z = \bar{s} \cdot a + s\cdot b$. PTM is described as a matrix of size $2\times8$ by considering select line $s$ in MSB and input $b$ in LSB.
\begin{equation}\label{eq13}\nonumber \footnotesize
V_{MUX}=
   \begin{bmatrix}
1 & 1 & 0 & 0 & 1 & 0 & 1 & 0 \\
0 & 0 & 1 & 1 & 0 & 1 & 0 & 1\\
\end{bmatrix}^\intercal
\end{equation}

\par Input matrix A given to the multiplexer is represented as 
$$
A=
\begin{bmatrix}\footnotesize
i_{000} &  i_{001} & i_{010} & i_{011} & i_{100} &  i_{101} & i_{110} & i_{111}\\
\end{bmatrix} $$

The output $Y$ using circuit PTM is represented as,
\begin{equation}\footnotesize\nonumber
    Y= A \times V_{MUX}
\end{equation}

\par\textit{\textbf{Definition 1}}: \textit{Scaled addition in the expression, $ Y_{sum}=\dfrac{p_a+p_b}{2} $ between inputs $A$, $B$ must be associated to select line $S$ of a multiplexer to satisfy the probability expression, $i_{001} + i_{110}= i_{010} + i_{101}$.}
\\
\par Consider three inputs $a$, $b$ and $s$ that are represented as Stochastic Number (SN) $A$,$B$ and $S$ respectively. The output of MUX \cite{gaines1967stochastic} is,
\begin{equation}\footnotesize
    Y_{sum}=\dfrac{p_a+p_b}{2}
\label{eqn:ysumsmall}
\end{equation}
 Using truth table representation, the output of MUX is 
\begin{equation}\footnotesize
    Y=i_{010}+i_{011}+i_{101}+i_{111}
\label{eqn:muxsop}
\end{equation}
\par The probabilities of each bitstream $A$,$B$, $S$ is given by the number of 1's occurring in each bitstream. Thus, 
\begin{equation}\footnotesize
\begin{aligned}
 p_{a}=i_{010}+i_{011}+i_{110}+i_{111}  \\ p_{b}=i_{001}+i_{011}+i_{101}+i_{111} \\ 
p_{s}=i_{100}+i_{101}+i_{110}+i_{111} 
\end{aligned} \Bigg \}
\label{eqn:pa_pb_ps}
\end{equation}
\par  Substituting $p_{a}$ and $p_{b}$ from Eq. \ref{eqn:pa_pb_ps} into Eq. \ref{eqn:ysumsmall}.
\begin{equation}\footnotesize
\label{eqn:ysum_new}
    Y_{sum}=\dfrac{i_{001}+i_{010}}{2} + i_{011} + \dfrac{i_{101} + i_{110}}{2} + i_{111}
\end{equation}
   \par Comparing Eq. \ref{eqn:ysum_new} and Eq. \ref{eqn:muxsop} we obtain
\begin{equation}\footnotesize
\label{eqn:muxc}
     i_{001} + i_{110}= i_{010} + i_{101};
     \end{equation} 
Eq. \ref{eqn:muxc} dictates the accurate condition of addition that eventuate of an association between two inputs and one select line. 
\vspace{2mm}
   \par  \textit{Example 1}: SN $A$ with $p_a=\frac{3}{8}$  as $10000011$ and $B$ with $p_b=\frac{5}{8}$ as $01111100$ are inputs to the MUX with $p_s=\frac{4}{8}$ as $01100011$ at select line $S$. The output is $11100000$ ($\frac{3}{8}$), whereas the expected output is $p_y=\frac{4}{8}$. This fallacy in output occurs as the probability of selection of input combinations is $(p_{010}+p_{101})=3\neq (p_{110}+p_{001})=5$ violating Eq. \ref{eqn:muxc}.
   \par However, keeping the input bitstream same if $p_s=\frac{4}{8}$ is changed to  $01100110$ the output bit
    stream becomes $11100010$($\frac{4}{8}$) which satisfies $p_{010}+p_{101}=p_{001}+p_{110}$.
    \begin{figure}[htp]
       \centering
       \includegraphics[height=2.4cm]{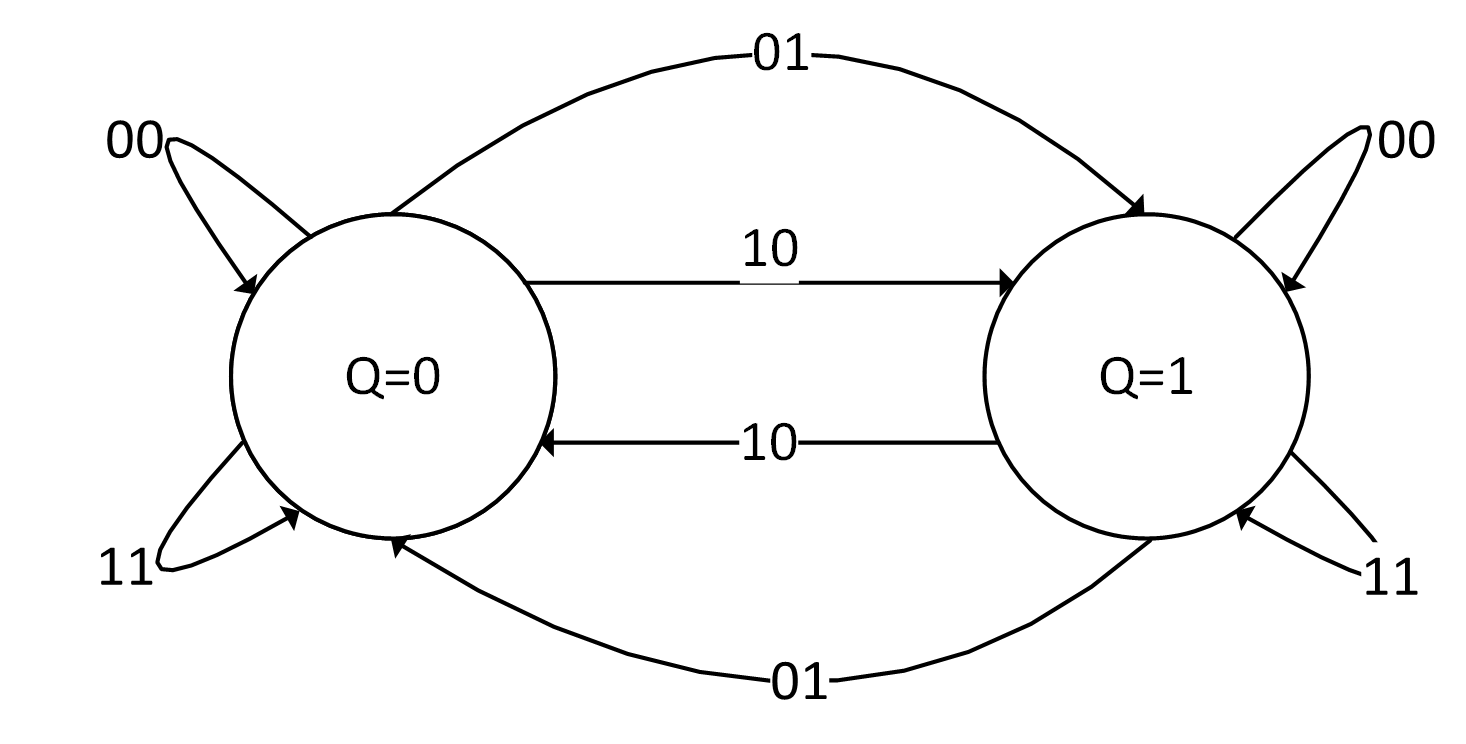}
       \caption{State diagram of adder validating Eq.\ref{eqn:muxc}.}
       \label{fig:state_diagram}
   \end{figure} 
   \par  Thus, there exists a strong dependence of the input bit streams on select line $S$ that must be satisfied  to have accurate addition. The state machine (see Fig. \ref{fig:state_diagram}) satisfying the condition is observed to behave accurately in presence of varying degrees of correlation between SNs. The state variable and the output of the FSM are the same and denoted by Q. It has been reported by Gaines \cite{gaines1967stochastic} that a T flip-flop gives a constant value of $0.5$ at the output irrespective of the input probability.
However, the additional constraint that had to be satisfied is fulfilled by using the XOR gate.  
   
   
\par b) \textit{Deriving condition for accurate Stochastic Subtraction:} A similar approach using the PTM can be adopted to analyze the behaviour of MUX to implement scaled subtraction \cite{alaghi2013exploiting}.  
\par\textit{\textbf{Definition 2}}: \textit{Scaled subtraction in the expression, $Y_{sub}$=$\dfrac{p_a+1-p_b}{2}$ between inputs $A$, $B$ must be associated to select line $S$ of a multiplexer to satisfy the probability expression,  $\dfrac{i_{100}+i_{000}}{2} + i_{110}= i_{101} + \dfrac{i_{011}+i_{111}}{2}$.}
\par Consider three inputs $A$, $B$ and $S$. The expression for scaled subtraction is given by  $Y_{sub}=\dfrac{p_a+(1-p_b)}{2}$. The probability of finding a '$0$' in $p_{b}$ can be given by $(1-p_{b})$, as \begin{equation}\footnotesize
\label{eqn:p_b0}
   (1-p_{b})=i_{000}+i_{010}+i_{100}+i_{101}
\end{equation}
\par Substituting $p_{a}$ and $p_{b}$ from Eq. \ref{eqn:p_b0} in $Y_{sub}$ we obtain,
\begin{equation}\footnotesize
\label{eqn:ysub}
    Y_{sub}=\dfrac{i_{000}+i_{100}}{2} + i_{010} + \dfrac{i_{011} + i_{111}}{2} + i_{110}
\end{equation}
   \par  Comparing Eq. \ref{eqn:ysub} and Eq. \ref{eqn:muxsop} we have,
\begin{equation}\footnotesize 
\label{eqn:mux_c}
     \dfrac{i_{100}+i_{000}}{2} + i_{110}= i_{101} + \dfrac{i_{011}+i_{111}}{2}
\end{equation} 
which is the condition for accurate subtraction involving MUX. To validate the appropriateness of the condition stated, two examples are cited.
\vspace{2mm}
\par \textit{Example $2$:} Consider $A$ and $B$ with $p_a=\frac{5}{8}$  as $11111000$ and $p_b=\frac{4}{8}$ as $10100101$ and the select input $S$ as $00111001$. The output is $01111010$ ($\frac{5}{8}$) which deviates from the expected result i.e., $\frac{9}{16}$. The error is much less in this uncorrelated number. Combinations of $A$, $B$ and $S$ shows that $N_{011}=1$,  $N_{010}=1$,  $N_{111}=1$,  $N_{110}=2$,  $N_{001}=1$,  $N_{000}=1$ and  $N_{101}=1$. Thus the condition given in Definition $2$ is violated. 
\par Now, if we consider correlated numbers the error is increased. Let SN A represented  as  $11111000$ ($p_a=\frac{5}{8}$) and B  as $11100000$ ($p_b=\frac{3}{8}$) and $S$ as $11110000$, then $Z$ is calculated as $11111111$ ($p_z=1$), whereas the expected output was $\frac{5}{8}$.
The analysis of the bitstreams with $S$ as MSB and $B$ as LSB shows that   $N_{010}=1$,  $N_{111}=3$,  $N_{110}=1$, and $N_{000}=3$. Placing values in the equation we find $\frac{3}{2}+1 \ne 0+\frac{3}{2}$.
 \vspace{1mm}
 \par Input stochastic signals fed to a combinational circuit can also be represented via PTM. We define an input vector of size $1 \times 2^k  $, where $'k'$ is the total number of input signals which when multiplied by the overall circuit PTM $M_{ckt}$ gives the output PTM. For a combinational circuit with uncorrelated inputs $X$ and $Y$ having probabilities $p_x$ and $p_y$ and output signal Z having probability $p_z$, we can write,
 \begin{equation}\footnotesize\nonumber
   I_{in} =
\begin{bmatrix}\footnotesize
    (1-p_x) &  p_x     
\end{bmatrix}
\otimes
\begin{bmatrix}\footnotesize
    (1-p_y)  &  p_y
\end{bmatrix} 
\end{equation}
\begin{equation}\footnotesize\label{eqn:Z}
   Z =
\begin{bmatrix}
    (1-p_z) &  p_z    
\end{bmatrix}
= I_{in}.M_{ckt}
\end{equation}
$I_{in}$ can also be written as
\begin{equation}\footnotesize\nonumber
  I_{in} =
\begin{bmatrix}
    i_0 &  i_1 & i_2 &  i_3 
\end{bmatrix} 
=
\begin{bmatrix}
    n_{00} &  n_{01} & n_{10} &  n_{11}
\end{bmatrix} 
\end{equation}where $i_0,i_1,i_2,i_3$ represents the probability of input bits $xy$ being $ 00,01,10$ and $11$ respectively. Thus, it can represent a correlation between input signals as well. For two maximally, minimally and  uncorrelated inputs, we can write \cite{alaghi2019accuracy}
\begin{equation}\footnotesize\label{eqn:I+1}
  I_{+1} = \begin{bmatrix}
    (1-p_x) &  0 &  (p_x - p_y) &  p_y 
    \end{bmatrix} ,\,\,p_x > p_y
    \end{equation}
 \begin{equation}\footnotesize\nonumber
   = \begin{bmatrix}
    (1-p_y) &  (p_y - p_x) & 0 &  p_x
    \end{bmatrix} ,\,\,p_y > p_x
\end{equation}
\begin{equation}\footnotesize\label{eqn:I-1}
  I_{-1} = \begin{bmatrix}
    \{1-(p_x+p_y)\} &  p_y &  p_x &  0
    \end{bmatrix} ,\,\,p_x +p_y \leq 1 
    \end{equation}
 \begin{equation}\footnotesize\nonumber
   = \begin{bmatrix}
    0 &  (1 - p_x) & (1 - p_y) & ((p_x+p_y)-1)
    \end{bmatrix} ,\,\,otherwise
\end{equation}
\begin{equation}\footnotesize\label{eqn:Io}
  I_0 = \begin{bmatrix}
    (1-p_x)(1-p_y) &  (1-p_x)p_y &  p_x(1-p_y) & p_xp_y 
    \end{bmatrix} 
    \end{equation}
With the help of Eqs. \ref{eqn:I+1}-\ref{eqn:Io} we can write the input vector matrix $I_{SCC}$ for any value of SCC.
\subsection{Reliability measure for circuits with varying correlation}
\par We are often concerned with the reliability of circuits under noisy conditions. The reliability of a circuit is defined as its ability to produce a correct output on a regular basis. For stochastic circuits, it can be evaluated using the circuit's ITM(J) and PTM(M). It can be shown that the reliability  of the circuit does not change with changes in  correlation, rather depends on the probabilistic error in the circuit.
\par \textit{\textbf{Definition 3}: The reliability of a circuit $R_{ckt}$ is invariant to change in correlation between  inputs and depends on the  probabilistic error rate '$p_e$'.}

Circuit reliability\cite{krishnaswamy2005accurate} is a measure of the similarity between it's ITM and PTM and is written as:
\begin{equation}\footnotesize
R_{ckt} = \sum_{J(i,j)=1}p(j|i).p(i) 
\label{eqn:RCKT_0}
\end{equation}
where, $p(j|i)$ is the $(i,j)^{th}$ entry of PTM. Using Eq. \ref{eqn:RCKT_0}, reliability of the circuit at  $SCC=0$ is obtained as ($1-p_e$). For AND gate, $R_{ckt}$ for different ranges of $SCC$ is obtained similarly  as:
\begin{equation}\footnotesize
R_{ckt}=
\begin{cases}
(p_e-1)(SCC(1-p_x-p_y +p_x.p_y)+p_xp_y-1)+\\(SCC(p_x+ p_y-1)-p_x.p_y(SCC+1))(p_e-1)\\
= 1-p_e, \hspace{1 cm}  \forall \hspace{2 mm} SCC < 0 \vspace{1 mm}\\
(p_e-1)(SCC.p_x+p_x.p_y-SCC.p_x.p_y-1)-\\(SCC.p_x-p_x.p_y.(SCC-1))(p_e-1 )
\\ = 1-p_e, \hspace{1 cm} \forall \hspace{2 mm} SCC >0
\end{cases}
\label{eqn:RCKT_SCC_0_1}
\end{equation}

\par Thus from Eq. \ref{eqn:RCKT_SCC_0_1}, it is observed that reliability of the circuit is independent of changes in correlation between the input numbers. Thus, error minimization by varying correlation does not affect the reliability of the circuit. This property is helpful in subsequent treatment of SLE to yield a correct output.


\begin{figure*}
     \centering
     \begin{subfigure}[b]{0.2453\textwidth}
         \centering
         \includegraphics[width=\textwidth]{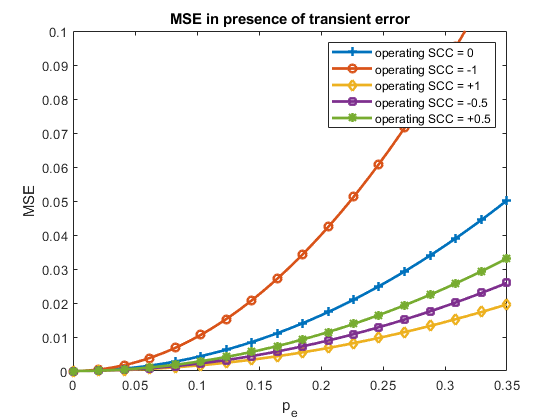}
         \caption{}
     \end{subfigure}
     \hfill
     \begin{subfigure}[b]{0.2453\textwidth}
         \centering
         \includegraphics[width=\textwidth]{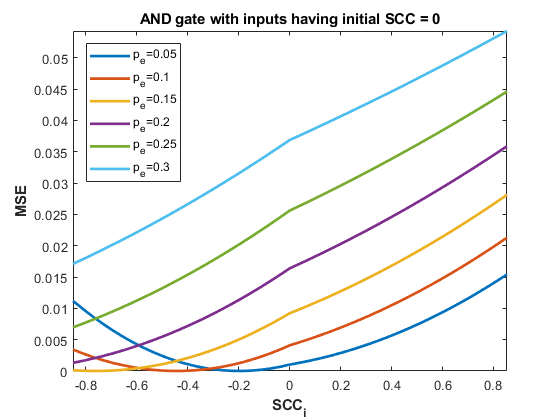}
         \caption{}
     \end{subfigure}
     \hfill
      \begin{subfigure}[b]{0.2453\textwidth}
         \centering
         \includegraphics[width=\textwidth]{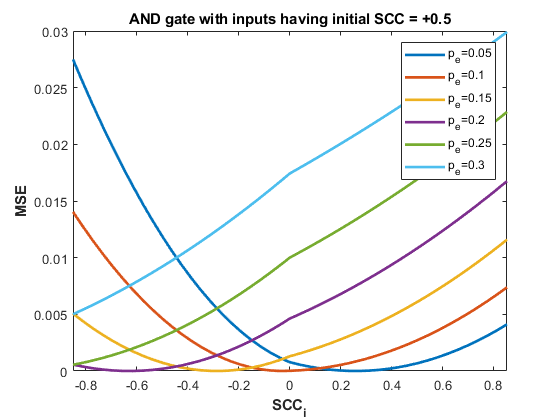}
         \caption{}
     \end{subfigure}
     \hfill
     \begin{subfigure}[b]{0.2453\textwidth}
         \centering
         \includegraphics[width=\textwidth]{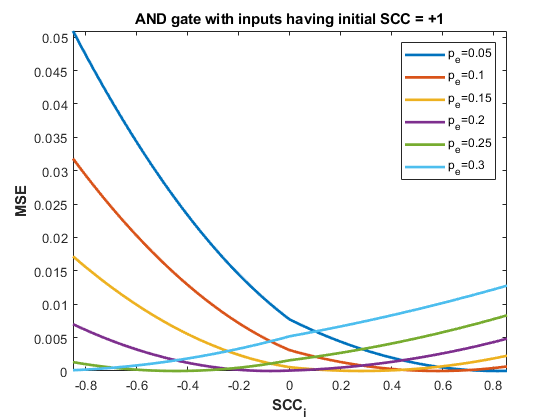}
         \caption{}
     \end{subfigure}
      \caption {MSE of AND gate  with varying transient errors; Min. MSE with ReCo (b) at $SCC=0$ (c) at $SCC=0.5$ (d) at $SCC=1$. }
      \label{fig:MSE_AND_GATE_graphs}
\end{figure*}

\begin{figure*}
     \centering
     \begin{subfigure}[b]{0.2453\textwidth}
         \centering
         \includegraphics[width=\textwidth]{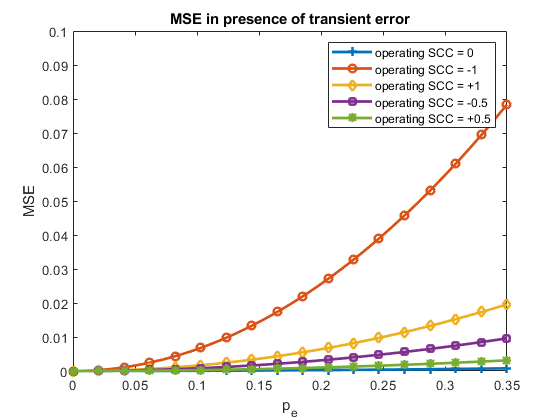}
         \caption{}
     \end{subfigure}
     \hfill
     \begin{subfigure}[b]{0.2453\textwidth}
         \centering
         \includegraphics[width=\textwidth]{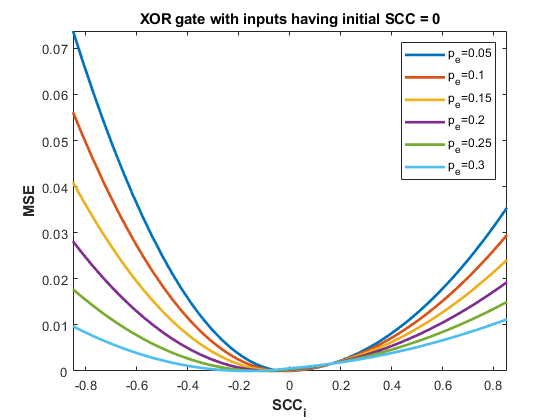}
         \caption{}
     \end{subfigure}
     \hfill
      \begin{subfigure}[b]{0.2453\textwidth}
         \centering
         \includegraphics[width=\textwidth]{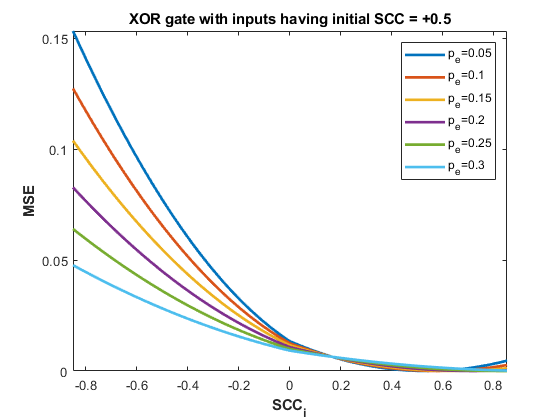}
         \caption{}
     \end{subfigure}
     \hfill
     \begin{subfigure}[b]{0.2453\textwidth}
         \centering
         \includegraphics[width=\textwidth]{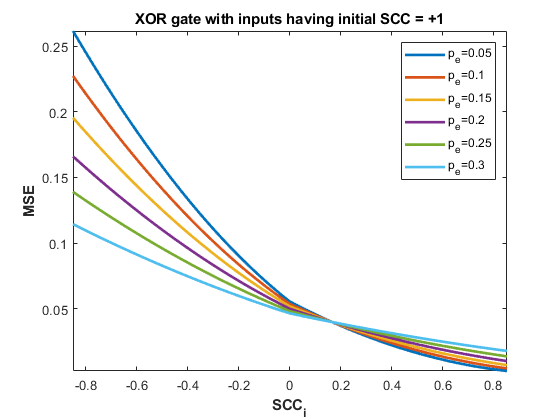}
         \caption{}
     \end{subfigure}
        \caption{ MSE of XOR gate  with varying transient errors; Min. MSE with ReCo (b) at $SCC=0$ (c) at $SCC=0.5$ (d) at $SCC=1$.}
        \label{fig:MSE_XOR_GATE_graphs}
        \end{figure*}
\section{Handling Errors in Stochastic Circuits using the Proposed technique}
\subsection{ Major Error Sources}
\subsubsection{Correlation Errors}
Correlation between two bitstreams has been identified as a major source of inaccuracy in certain stochastic circuits.
Correlation in Stochastic computing indicates that the bitstreams generated by LSFR's \cite{anderson2016effect} or  SNGs \cite{alaghi2013exploiting} inherit some sort of dependence between them (cross correlation) or between the bits of the same  bitstream (auto correlation). Earlier, correlation in stochastic circuits could only have been vaguely identified as inaccurate output caused by a pair of bitstream when passed through an AND gate. But, only recently correlation in stochastic computing has been quantified and identified with definiteness \cite{alaghi2013exploiting}.
\par
\begin{figure}[htp]
    \centering
    \includegraphics[width=4.6cm]{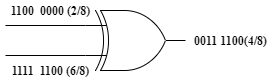}
    \caption{XOR gate as absolute subtractor when inputs are  positively correlated. }
    \label{fig:xor}
\end{figure}

 To quantify the correlation between input bitstreams $X$ and $Y$, $SCC$ (Stochastic Correlation Coefficient) which is analogous to the similarity coefficient \cite{ahlgren2003requirements} is represented as
\begin{equation} \nonumber \footnotesize
SCC(X,Y)=
\begin{cases}
\frac{p_{X \land Y}-p_X\cdot p_Y}{min(p_X,p_Y)-p_Xp_Y} \text{ , } p_{X \land Y}> p_X\cdot p_Y\\
\frac{p_{X \land Y}-p_X\cdot p_Y}{p_Xp_Y-max(p_X+p_Y-1,0)} \text{ , otherwise } 
\end{cases}
\end{equation}
where, $p_{x \land y}$ is obtained by bitwise AND operation between $X$ and $Y$.
Other generalized way of representing the SCC is:
\begin{equation} \nonumber \footnotesize
SCC(X,Y)=
\begin{cases}
\frac{n_{11}.n_{00}-n_{01}.n_{10}}{n.min(n_{11}+n_{10},n_{11}+n_{01})-(n_{11}+n_{10})(n_{11}+n_{01})}\\  \qquad \qquad  \qquad  \qquad \text{,} \quad    n_{11}.n_{00}  > n_{01}.n_{10}\\

 \frac{n_{11}.n_{00}-n_{01}.n_{10}}{(n_{11}+n_{10})(n_{11}+n_{01})-n.max(n_{11}-n_{00},0)}\\ \qquad \qquad  \qquad  \qquad \text{ ,otherwise }    
\end{cases}
\end {equation}
where, $n_{11}$, $n_{10}$, $n_{01}$ and $n_{00}$ are respective overlaps of $X$ and $Y$. Thus, the measure of correlation is influenced only by the overlap of similar and dissimilar bits in both the bitstreams. Let, $X = 1100 1111 0100$, and $Y = 0100 1111 0100$,  then, $SCC = +1$. But, if $X = 1011 0001 0101$, $Y =1111 1100 0101$, $SCC = 0.5$. In this case, not every $1$ in Y is influenced by the presence of 1 in that particular position in $X$. There is overlapping of $0's$ in $X$ and $1's$ in $Y$ as well as $1's$ in $X$ and $0's$ in $Y$. Thus, the pair of bitstream is positively correlated to certain degree.
 Correlation has also been found to impact the circuit's behaviour in a positive way \cite{budhwani2017taking}. 
XOR gate acts as an absolute subtractor when inputs are positively correlated as shown in Fig. \ref{fig:xor}. Implementing the same function using binary inputs increased hardware complexity \cite{kanoh1990absolute}.
\par But for boundary  values of probability, either, $0$ or $1$, the measure of SCC becomes indeterminate.
 To relate to this, consider two SNs $X = 0000 0000$ and $Y = 1111 1111$. Logic operations on these numbers will produce output that will stick to the boundary values itself, either $0$ or $1$ depending on the SLE. Attempts to change the correlation status will result in a change in probability value which is undesired.
Using a correlator circuit such as \cite{lee2018correlation} will not be able to alter the degree of correlation between $X$ and $Y$ because only grouping of one kind of bit-pair (here $01$) will be possible and we lose the leverage of pairing other three bit pairs. 
\vspace{5mm}
For any degree of correlation, we can write the output $p_z$ as a linear equation\cite{alaghi2013exploiting}, given as:
\begin{equation}\label{eqn:pzSCC<0} \footnotesize
p_z=(1+\,SCC)F_0 \,- SCC.F_{-1}, \;\; \forall \; SCC<0
\end{equation}
\begin{equation}\label{eqn:pzSCC} \footnotesize
p_z=(1-\,SCC)F_0 \,+ SCC.F_{+1} , \;\; \forall \; SCC>0
\end{equation}
$F_0,F_{-1},F{+1}$ are the functions realized by the logic gate with SCC values of 0,-1,+1.
\vspace{2mm}
\begin{figure*}
     \centering
     \begin{subfigure}[b]{0.32\textwidth}
         \centering
         \includegraphics[width=\textwidth]{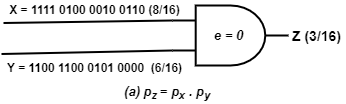}
         \label{SOFT ERROR 1}
    
     \end{subfigure}
     \hfill
     \begin{subfigure}[b]{0.32\textwidth}
         \centering
         \includegraphics[width=\textwidth]{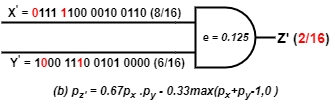}
         \label{SOFT ERROR 2}
        
     \end{subfigure}
     \hfill
      \begin{subfigure}[b]{0.32\textwidth}
         \centering
         \includegraphics[width=\textwidth]{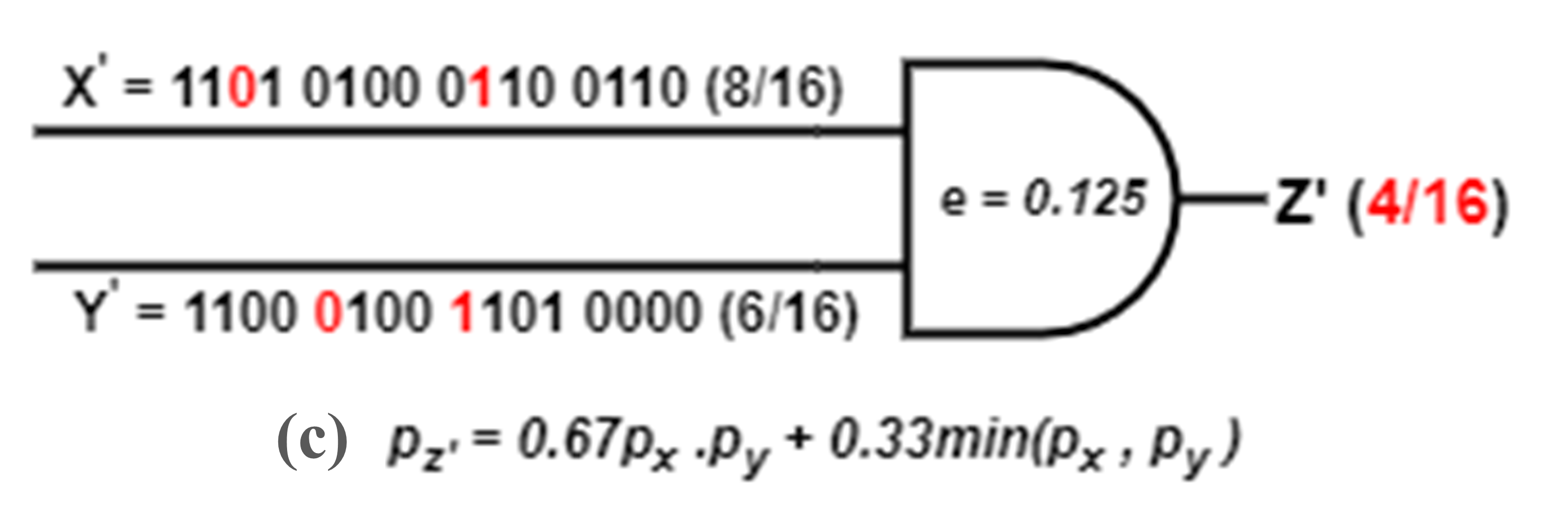}
         \label{SOFT ERROR 3}
      
     \end{subfigure}
     \caption{Bit flips at different positions due to transient errors and it's impact on correlation alteration resulting in different SFs.}
     \label{fig:Different_s.f}
\end{figure*}

\subsubsection{Soft Errors}\label{AA}

	As semiconductor technology advances with reduced feature size and increased scalability, it is becoming more prone to soft errors  \cite{nicolaidis2005design}. The sources of such soft errors have been traced to mainly alpha particles and high energy cosmic rays \cite{shivakumar2002modeling}. Although soft errors are not unique to stochastic circuits, its properties make it more tolerant to soft errors than weighted-binary logic circuits. Soft errors do not affect the circuit physically but it introduces behavioral changes in the circuit in the form of bit flips by introducing false logic\cite{alaghi2017trading} \cite{chen2014behavior}.
	Transient or soft errors are caused due to exposure to external radiation  and are greatly increased by manufacturing defects in a chip. These introduce false logic at the output of the circuit \cite{nicolaidis2005design}. Thus, if multiple faults strike nodes of a gate, the output may be obtained erroneously.
	Bit flips are modelled as bit flip error $p_e$ associated with each gate in the circuit as a Bernoulli variable. 
	\par Stochastic numbers are analyzed as Bernoulli random variables (BRV) represented by its probability of success $p_x$ to perform similar operations as with other BRVs. Errors in  BRV's are usually analysed using Mean Square Error (MSE) which is written as $ E_z=E[(p_{ze}-p_z)^2]$, where, $p_{ze}$ and $p_z$ represent the estimated and exact value respectively.
	A lot of applications involving stochastic circuits are carried out in a noisy environment where the circuit is prone to bit-flip errors. For nano-scale devices transient or soft errors are growing prominence as the device features are downscaled to sub-micron ranges. The observed output might exceed the error threshold  due to the change in the expected value of signals and also due to unwanted correlation introduced during bit flips. For larger circuits this may be a major concern for accuracy \cite{chen2014analyzing}.
	\par The presence of soft errors coupled with other inherent error sources may cause models instability in multiple responses.
	Thus to achieve the desired level of accuracy irrespective of the environment is a dire need in this scenario. Soft errors can change the status of correlation between bitstreams by changing the probability value of the inputs. If equal number of 1s and 0s are flipped on a bitstream on account of transient faults, then the probability value remains unchanged. Fig. \ref{fig:Different_s.f} shows the effect of transient errors on the behaviour of an AND gate. At zero fault rate, the AND gate implements exact multiplication of two numbers as shown in Fig. \ref{fig:Different_s.f} (a).
	This condition is not true for two other cases, when, $p_e$ at $0.125$ hit upon the input nodes at different bit positions leading to varied correlation status between two numbers as shown in Fig. \ref{fig:Different_s.f} (b) and (c).
	\par As we increase the amount of transient errors in the circuit the MSE increases exponentially. In case of inputs operating in the negative range of correlation the error surmounts with the incremental injection of soft error rates as shown in Fig. \ref{fig:MSE_AND_GATE_graphs} (a) (red color) when compared to the same bitsreams operating in the positive range of correlation  showing reduced MSE with the injection of soft errors (yellow color). These are discussed  in detail in the next section, where the motto is to reduce the effect the transient faults on probabilistic circuits by harnessing some of the unique properties of each of these correlation sensitive circuits.
\subsection {The proposed Remodelling Correlation (\textit{ReCo}) Framework}
 Minimizing errors is crucial since this distorts output logic level of the circuit. We assume that transient faults at the gates caused due to environmental conditions lead to change in the input as well as output
 probabilities thereby introducing uncertainty in correlation assumption of the circuit. It is observed that bit flips at different positions due to transient errors may lead to different correlation status between the same bitstreams. Undesired correlation can also lead to different stochastic functions being implemented by the same logic circuit as shown in Fig. \ref{fig:Different_s.f} and impedes the natural function to get implemented. Change in correlation status may also result in the change in probability value if an unequal number of $0's$ and $1's$ are flipped.
 \par
\begin{figure}
     \centering 
     \includegraphics[ width=0.40 \textwidth]{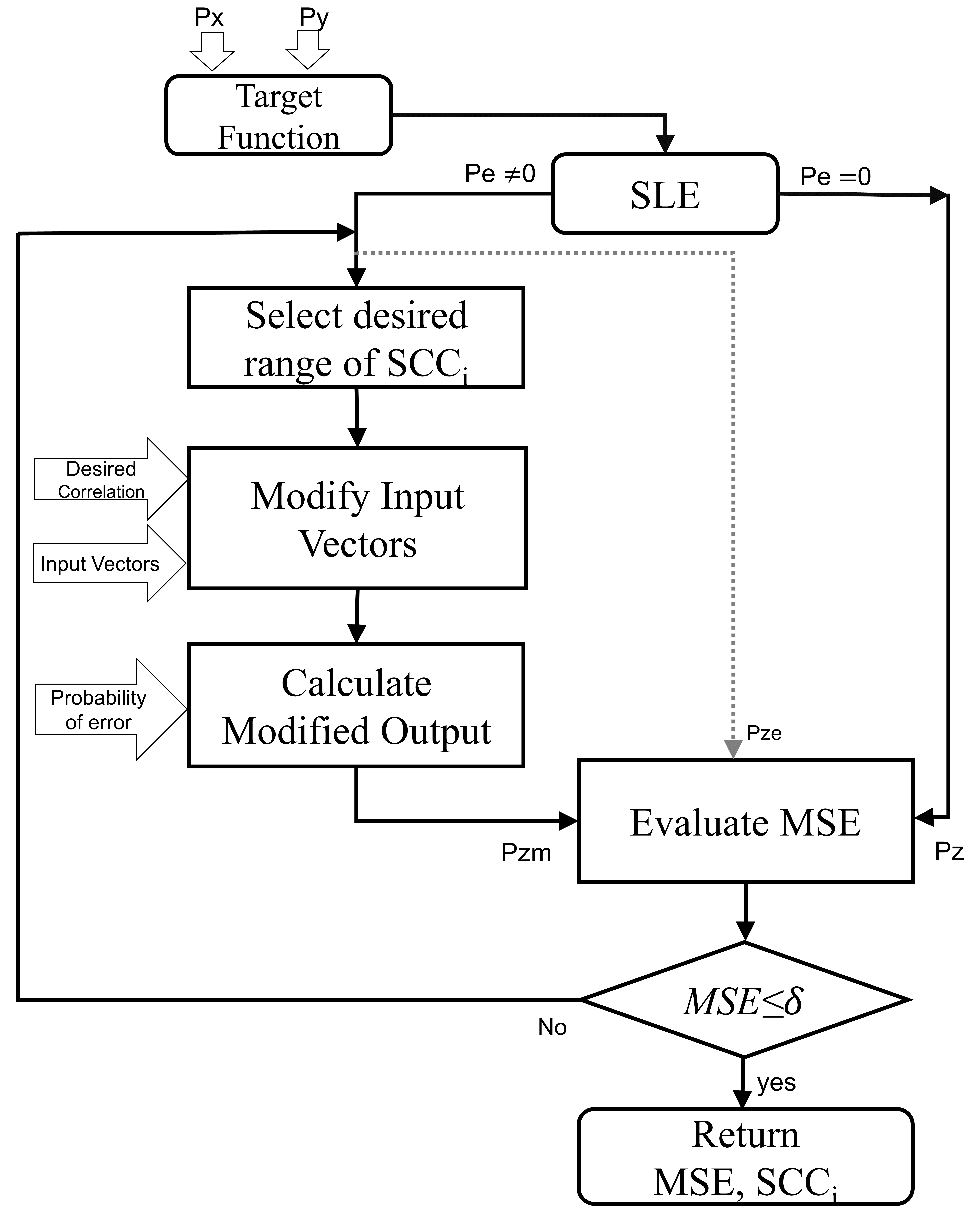}
     \caption{The Flowchart of the Proposed Framework.}
     \label{fig:Reco_SLE}
 \end{figure}

 Our work suggests \textit{Remodelling Correlation (\textit \textit{ReCo})} technique to cater to the change in the probability assumption at the inputs owing to transient faults. We interpret techniques for correlation-sensitive elements to bring down the MSE to a minimum level.  While conducting
a study on  correlation-sensitive SLEs we demonstrate that every design error can be corrected by introducing correlation to a certain degree at the inputs.  We deduce an operating point of the circuit in this incorrect environment with a suitable injection of SCC that reduces MSE to a minimum value. $Algorithm 1$ searches for a unique solution of the induced correlation within the range [\ -1,1]\ to find a minimum error for input parameters. We begin our analysis first by considering single SLEs. The flowchart of the proposed method is shown in Fig. \ref{fig:Reco_SLE}.
\subsubsection{  ReCo analysis for correlation sensitive elements with zero  correlation assumption}\textcolor{white}{.} A stochastic circuit implements different real-valued functions when the correlation status between input SNs are altered. We assume the target function to be implemented at $SCC(X,Y)=0$ and any deviation is considered as faulty behaviour of the circuit.
\par  i) AND gate: Consider an AND gate that is inflicted by transient noise. We assume $SCC(X,Y)=0$ , so $p_z=p_xp_y$. As we increase the probability of transient error the observed output deviates more from the original output and there is an exponential increase in MSE, indicated  in Fig. \ref{fig:MSE_AND_GATE_graphs}(a) (blue). So we attempt to reduce the MSE using the proposed method.
\par We first modify  input vectors between $p_x$ and $p_y$. Assuming $p_x$ $<$ $p_y$ and {$p_x$ + $p_y$ $\leq$ 1}, the modified vectors of $I_{SCC_{m}}$ are,
\begin{equation}\footnotesize\label{eqn:ISCCm_1}
 I_{SCC_{m}}= 
\footnotesize\begin{bmatrix}
   & 1-(p_x+p_y)+p_xp_y(1+SCC_i)
 \\ & -p_y(p_x+p_xSCC_i - 1)
 \\  & -p_x(p_y+p_ySCC_i - 1)
 \\ & p_xp_y(SCC_i+ 1)
    \end{bmatrix}^\intercal
\end {equation}
For $p_x$ + $p_y$ $>$ 1,
\begin{equation}\footnotesize\label{ISCCm_2}
 I_{SCC_{m}}= 
\footnotesize\begin{bmatrix}
   & -(p_y-1)(p_xSCC_i-p_x+1)
 \\ & p_y(SCC_i-1)(p_x-1)-SCC_i(p_x-p_y)
 \\  & p_x(SCC_i-1)(p_y-1)
 \\ & p_xSCC_i-p_xp_y(SCC_i-1)
    \end{bmatrix}^\intercal
\end {equation}

\begin{algorithm}[h]
\begin{algorithmic}[1]
\caption{ \textit{ ReCo analysis for a single gate}}
\STATE \textbf{Input} {$p_x,p_y,p_e,input\_Gate$}; \textbf{Output} {$MSE_i,SCC_i$}
\STATE  $\textit{\textbf{ReCo}}(input\_Gate)$
\STATE [$p_x,p_y$]= input probabilities of $input\_Gate$;
\STATE $p_e$= probabilities of transient errors of $input\_Gate$;
\STATE $True\_Output=\textbf{\textit{Eval}}(p_x,p_y,SCC);$
\STATE \For{$SCC_i=-1;SCC_i<= +1;SCC_i+=0.001$} 
{ 
$Modified\_Output=\textbf{\textit{Eval}}(p_x,p_y,p_e,SCC_i);$\\
$MSE_i=True\_Output-Modified\_Output$ \\

 \If{$MSE_i \leq \delta $}
  {
    return $MSE_i$ ,  $SCC_i$
  }
}
\STATE { return $\argmin_{MSE_i}$  \{$MSE_i$ ,  $SCC_i$\}}
\end{algorithmic}
\end{algorithm}

For an AND gate with a given  error rate $p_e$, we write modified output $p_{zm}$ as a function of  $I_{SCC_m}$.
\begin{equation}\footnotesize\label{eqn:pzm1}
    p_{zm}=I_{SCC_{m}}
 \times\begin{bmatrix}
    1-p_e &  p_e \\
    1-p_e  &  p_e \\
    1-p_e  & p_e \\
    p_e &  1-p_e \\
\end{bmatrix} 
\end{equation}

\begin{figure*}
     \centering
     \begin{subfigure}[b]{0.2453\textwidth}
         \centering
         \includegraphics[width=\textwidth]{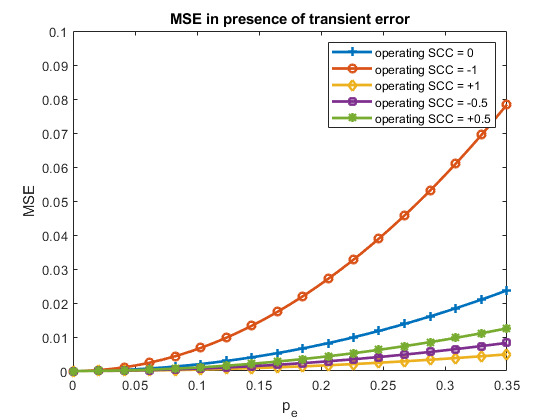}
         \caption{}
         \label{}
     \end{subfigure}
     \hfill
     \begin{subfigure}[b]{0.2453\textwidth}
         \centering
         \includegraphics[width=\textwidth]{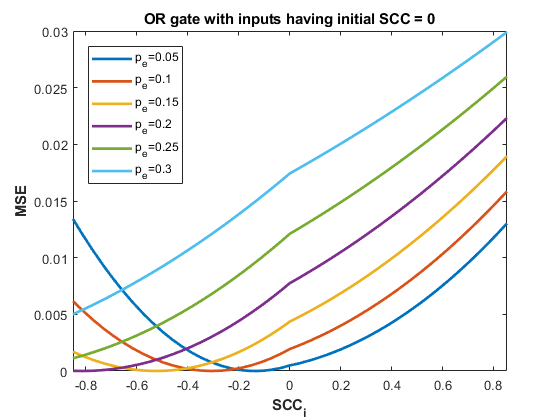}
         \caption{}
         \label{fig:OR2}
     \end{subfigure}
     \hfill
      \begin{subfigure}[b]{0.2453\textwidth}
         \centering
         \includegraphics[width=\textwidth]{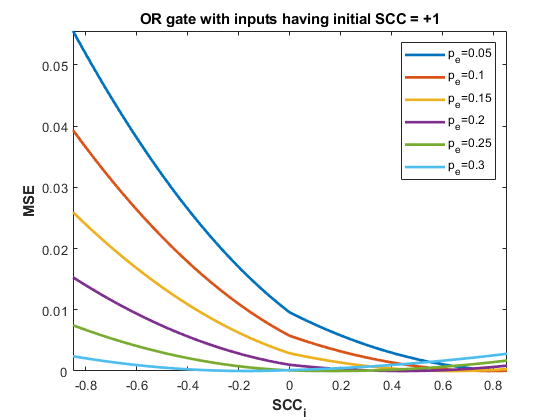}
         \caption{}
         \label{fig:OR3}
     \end{subfigure}
     \hfill
     \begin{subfigure}[b]{0.2453\textwidth}
         \centering
         \includegraphics[width=\textwidth]{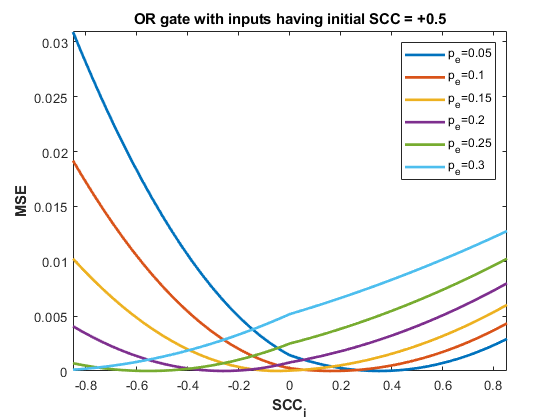}
         \caption{}
         \label{fig:OR4}
     \end{subfigure}
        \caption{(a) MSE of OR gate  with varying transient errors; Min. MSE with ReCo (b) at $SCC=0$ (c) at $SCC=1$ (d) at $SCC=0.5$.
        }
        \label{fig:MSE_OR gate}
\end{figure*}

$p_{zm}$ is observed as $f(SCC_i)$. We try to make $p_{zm}$ close to $p_z$ to reduce the observed error $p_{ze}$. Thus, 
\begin{equation}\footnotesize\label{eqn:pzm2}
\begin{multlined}
p_{zm}=p_e+p_xp_y(1+SCC_i)(1-2p_e)
\end{multlined}
\end{equation}
The MSE which is $(p_{zm}-p_z)^2$ is calculated as,
\begin{equation}\footnotesize\label{eqn:MSEand}
MSE_{and} = \{p_e+p_xp_y(SCC_i-2p_e- 2p_eSCC_i)\}^2
\end{equation}
The induced $SCC_i$ which reduces MSE to a minimum possible value within  the range [-1,+1] is obtained by differentiating Eq. \ref{eqn:MSEand} w.r.t $SCC$ and equate it to $0$.
\begin{equation}\footnotesize\nonumber
\begin{split}
 2p_xp_y(1-2p_e)(p_e+p_xp_y(SCC_i-2p_e-2p_eSCC_i))=0
\end{split}
\end{equation}
\begin{equation}\label{eqn:SCC_i_1}\footnotesize
 \therefore SCC_i = \frac{-(p_e-2p_ep_xp_y)}{p_xp_y(1-2p_e)}
\end{equation}
Eq. \ref{eqn:SCC_i_1} dictates the condition of reaching a minimum value of MSE for $SCC_i$ in the range [-1,0]. Similarly,   when $p_x + p_y >1$ with $p_x<p_y$, $SCC_i$ can be evaluated as,
\begin{equation}\label{eqn:SCC_i_2}\footnotesize
SCC_i =  \frac{(p_e-p_x-p_y+ p_xp_y-2p_ep_xp_y+1)}{(2p_e - 1)(p_x - 1)(p_y - 1)}
\end{equation}
where, $p_x,p_y>0 $ and $p_x<p_y$. Thus, $p_xp_y>0$.
Expressions are derived assuming negative induction of correlation. However, nothing in the derivation prevents $SCC_i$ from being positive to achieve minimum MSE.

\par To exploit the simplicity of equations and to achieve maximum possible accuracy in calculations, parameters appearing in equations are verified graphically. Fig. \ref{fig:MSE_AND_GATE_graphs}(b) shows different values of induced correlation to obtain zero error at the output. Note that, Eq. \ref{eqn:SCC_i_1},\ref{eqn:SCC_i_2} always hold for $p_e<0.5$. Using similar analysis we arrive to different sets of equations for $p_x$ $>$ $p_y$.
\par
\begin{figure*}
     \centering
     \begin{subfigure}[b]{0.31\textwidth}
         \centering         \includegraphics[width=\textwidth]{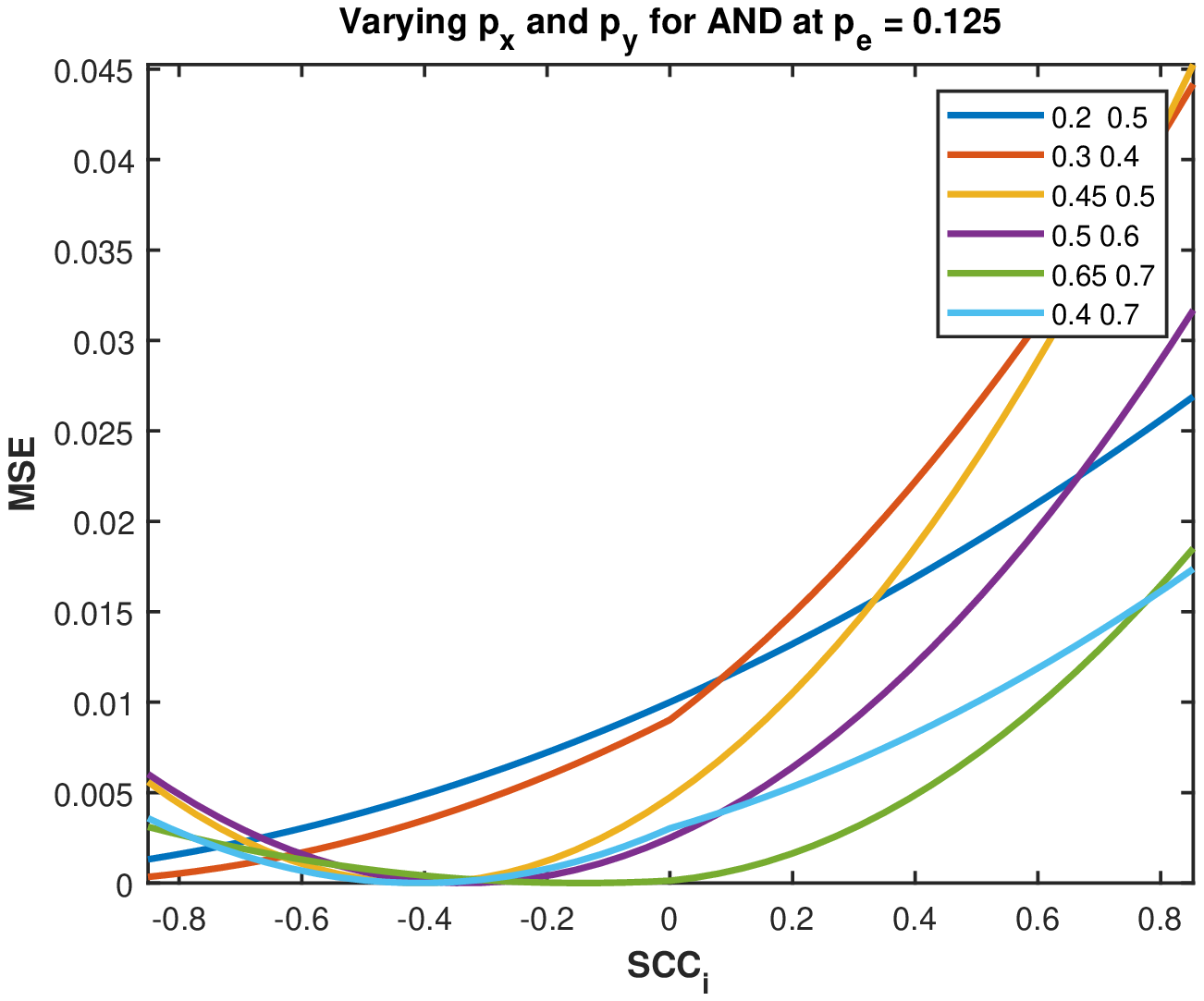}
         \caption{}
     \end{subfigure}
     \hfill
     \begin{subfigure}[b]{0.31\textwidth}
         \centering
         \includegraphics[width=\textwidth]{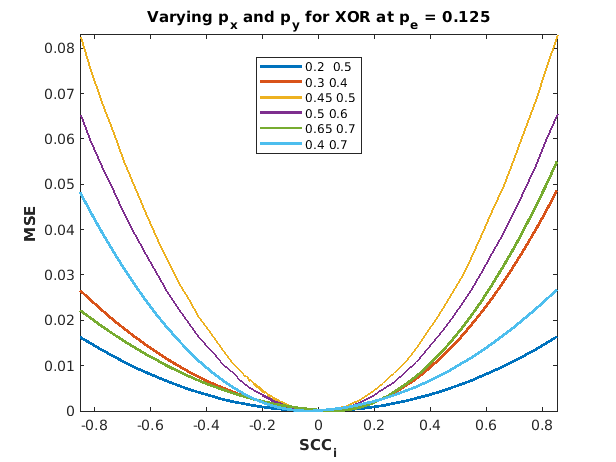}
         \caption{}
     \end{subfigure}
     \hfill
      \begin{subfigure}[b]{0.31\textwidth}
         \centering
         \includegraphics[width=\textwidth]{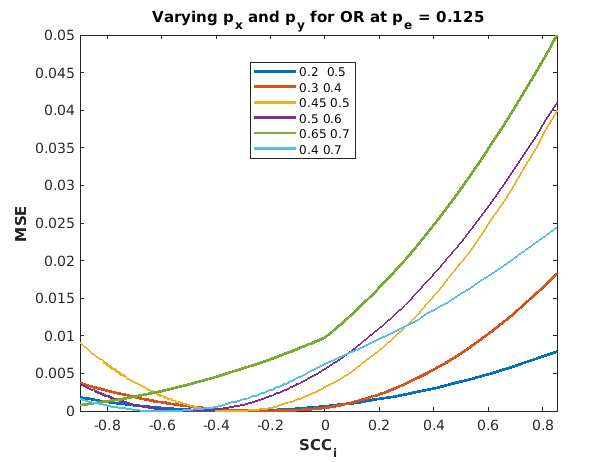}
         \caption{}
     \end{subfigure}
     
    \caption{MSE after ReCo analysis of correlation sensitive gates for varying input probabilities.}
    \label{fig:MSE_Varying_input_probabilities}
\end{figure*}

 \textit{Example $3$:} Consider an AND gate with $p_x=0.3$ and $p_y=0.6$. The error-free output is $p_z$ $=0.3\times0.6=0.18$. The observed output is $p_{ze}=0.28$ at $p_e=0.15$. Thus,
\begin{equation}\nonumber\footnotesize
  I_{SCC_{m}}= 
\footnotesize\begin{bmatrix}
   & (0.28+0.18SCC_i) \\ & (0
   .42-0.18SCC_i) \\  & (0.12-0.12SCC_i) \\ & (0.18+0.12SCC_i)
    \end{bmatrix}^\intercal  
\end{equation} 
Thus, $p_{zm}$=$0.18SCC_i + 0.64p_e - 0.36p_eSCC_i + 0.18$ and 
 $MSE$ = $\frac{(9SCC_i + 32p_e - 18p_eSCC_i)^2}{40000}$.
 Error reduces to $0$ for  $SCC_i=-0
 .76$.  It is observed that MSE can be reduced to $0$ if $p_e\leq 0.2$ (a considerate limit). The results are confirmed graphically considering different values of $p_x$ and $p_y$ at $p_e=0.125$  as shown in Fig. \ref{fig:MSE_Varying_input_probabilities}(a).
\par \textit{ii) XOR gate:} At $SCC=0$, the XOR gate implements $p_z=p_x(1-p_y)+p_y(1-p_x)$.  Any deviation from the target function on account of transient error is considered as a contribution to MSE. A similar foregoing approach is adopted in the analysis of the XOR gate to operate at minimum MSE under different transient error rates. 
Now, $p_{zm}$ for $p_x<p_y$ is written as, 
\begin{equation}\footnotesize\label{eqn:pzm3}
    p_{zm}=I_{SCC_{m}} \times\begin{bmatrix}
    1-p_e &  p_e \\
    p_e  &  1-p_e \\
    p_e  & 1-p_e \\
    1-p_e &  p_e \\
\end{bmatrix} 
\end{equation}

Substituting $I_{SCC_{m}}$ from Eq. \ref{eqn:ISCCm_1},
\begin{equation}\footnotesize\label{eq35}\nonumber
\begin{split}p_{zm} = p_e-p_xp_y-2p_e(p_x+p_y-2p_xp_y)-2p_xp_ySCC_i(1-2p_e)
 \end{split}
\end{equation}
\begin{equation}\footnotesize\label{eqn:MSE_xor}
\footnotesize \begin{split}
MSE_{xor} = \{2p_e(p_x+p_y)-p_e+2p_xp_y(SCC_i-2p_e-2p_eSCC_i)\}^2
\end{split}
\end{equation}
We differentiate Eq. \ref{eqn:MSE_xor} w.r.t $SCC_i$ and equate it to $0$.
\begin{equation}\footnotesize \nonumber
\begin{multlined}
 -2(2p_xp_y-4p_ep_xp_y)(p_e-2p_ep_x-2p_ep_y-\\2p_xp_y(SCC_i+2p_e+ 2p_eSCC_i)=0
\end{multlined}
\end{equation}
\begin{equation}\label{eqn:SCC_i_3}\footnotesize
\therefore SCC_i= \frac{p_e-2p_e(p_x+p_y)+4p_ep_xp_y}{2p_xp_y-4p_ep_xp_y}
\end{equation}
Similarly, for $p_x+p_y>1$,
\begin{equation}\footnotesize \label{eqn:SCC_i_4}
\begin{multlined}
SCC_i= \frac{p_e-2p_ep_x-2p_ep_y+4p_ep_xp_y}{4p_e +2(p_x+p_y)(1-2p_e)-2p_xp_y(1+2p_e)-2}\end{multlined}
 \end{equation}
\par Fig.  \ref{fig:MSE_XOR_GATE_graphs}(b) shows different values of $SCC_i$ to reach zero MSE at different values of $p_e$, which is consistent with Eq. \ref{eqn:SCC_i_3}.
\\\textit{Example 4:}  Consider XOR gate with inputs $p_x = 0.3$ and $p_y =0.6$. Thus $p_z=0.54$ and $p_{ze}=0.52$ at $p_e=0.15$. By substituting $I_{SCC_{m}}$ we find
$p_{zm} = 0.72p_e - 0.08p_e - 0.36SCC_i + 0.54$ and hence
$
MSE =\frac {(9SCC_i+2p_e-18p_eSCC_i)^2}{625} \nonumber
$. Thus, $MSE$ reduces to $0$ at $SCC_i=0$.
\par From the experiment (see Fig. \ref{fig:MSE_XOR_GATE_graphs} (a)) it is inferred that the XOR gate is least responsive to probabilistic errors and hence the smallest contributor to the overall MSE in the circuit. Also, it is evident from Fig. \ref{fig:MSE_XOR_GATE_graphs}  (b) that the XOR gate is most sensitive to changes in correlation. Thus for $0\leq p_e \leq 0.3 $, the error can be reduced to 0 in contrary to all other gates (see Fig.\ref{fig:MSE_AND_GATE_graphs}(b) and Fig. \ref{fig:MSE_OR gate}(b)).
This is also validated using different values of $p_x$ and $p_y$ at $p_e=0.125$ as shown in Fig. \ref{fig:MSE_Varying_input_probabilities}(b).
\vspace{2mm}
\par  \textit{iii) OR gate:} For uncorrelated numbers, OR gate implements $p_z=p_x+p_y-p_xp_y$. In presence of transient error let the function be $p_{ze}$. We eliminate this error by introducing the  \textit{ReCo} block at inputs to inject the desired correlation. The modified output $p_{zm}$ is then calculated using  Eq. \ref{eqn:ISCCm_1}.
\begin{equation}\footnotesize\label{eq41}
p_{zm}=I_{SCC_{m}} \times\begin{bmatrix}
    1-p_e &  p_e \\
    p_e  &  1-p_e \\
    p_e  & 1-p_e \\
    p_e &  1-p_e \\
\end{bmatrix} 
\nonumber
\end{equation}
\begin{equation}\footnotesize\label{eq42}\nonumber
\begin{split} 
\therefore p_{zm} = p_e+p_x+p_y-2p_ep_x-2p_ep_y-p_xp_y-\\p_xp_ySCC_i+2p_ep_xp_y+2p_ep_xp_ySCC_i
\end{split}
\end{equation}
\begin{equation}\label{eqn:MSE_or}\footnotesize
\begin{split}
\therefore MSE_{or}=(p_e\{1+p_xp_y\}-2p_e\{p_x+p_y\}-p_xp_ySCC_i\{1-2p_e\})^2  
\end{split}
\end{equation}
We differentiate Eq. \ref{eqn:MSE_or} w.r.t $SCC_i$ and equate it to $0$.
\begin{equation}\footnotesize\label{eq43}
\begin{split}\nonumber
-2(p_xp_y-2p_ep_xp_y)(p_e-2p_ep_x-2p_ep_y\\-p_xp_ySCC_i+2p_ep_xp_y+ 2p_ep_xp_ySCC_i)=0
\end{split}
\end{equation}
\begin{equation}\label{eqn:SCC_i_5}\footnotesize
\therefore SCC_i = \frac{p_e-2p_e(p_x+p_y)+2p_ep_xp_y}{p_xp_y-2p_ep_xp_y}
\end{equation}
Similarly for $p_x+p_y>1$,
\begin{equation}\label{SCC_i_6} \footnotesize
SCC_i =\frac{2p_e(p_y-p_x)-p_e+p_xp_y(1-2p_e)}{p_xp_y(1-2p_e)}
\end{equation}
Fig.  \ref{fig:MSE_OR gate}(b) shows different values of induced correlation to obtain zero error at the output for different values of $p_e$.
\par \textit{Example $5$:} Consider an OR gate with inputs $p_x=0.3$ and $p_y =0.6$ at $p_e = 0.15$. The error free output $p_z=0.72$ and the observed output $p_{ze}=0.654$.
Using Eq. \ref{eq43} 
$
p_{zm} = 0.36p_eSCC_i-0.44p_e-0.18SCC_i+ 0.72 $
and 
$MSE =\frac {(9SCC_i+22p_e-18p_eSCC_i- 6)^2}{2500}$. Thus, MSE can be reduced to zero at $SCC_i = -0.5238$.
\par Thus similar to AND gate, the MSE of the OR gate can be reduced to 0 if the probabilistic error value $p_e$ is below a certain limit i.e $0.2$, but the amount of reduction that is achieved is less compared to AND gate.
\par From the analysis it is observed that the OR gate is least sensitive to changes in correlation, whereas  XOR stands  highest in the sensitivity list. AND gate is intermediate to them. It is also identified that the XOR gate is least affected by the transient error whereas AND gate is mostly influenced by the presence of transient error. So MSE increases immensely when the error is imposed on an AND gate. These properties of the XOR gate make it a suitable choice for the analysis of an error-resilient circuit design. In the next section, this idea is implemented on complex circuits that focus to minimize MSE using the minimum hardware in the correction circuit
using the proposed analysis.
\subsubsection {\textit \textit{ \textit{      ReCo}} Analysis for correlation-sensitive logic elements with non-zero correlation assumption} 
Those SLEs which are sensitive to correlation implement an altogether different stochastic function as in the case of AND gate shown in Fig.  \ref{fig:Different_s.f}(b),(c). In this section, we  reconsider SLEs with transient errors having an apriori correlation assumption. We invoke \textit \textit {ReCo} analysis to suppress MSE and formulate the underlying conditions in support of that. Two distinct cases of initial correlation assumption are discussed i.e., $SCC=+0.5$ and $SCC=+1$ and perform a similar analysis to reduce errors at different degrees of transient faults. The target function is
obtained considering an initial non-zero and positive value of correlation. It is observed that for an existing negative SCC between input variables the effect of transient errors in the circuit element is enhanced. Thus such cases are excluded in our analysis.
\par i)\textit{ AND gate:}
 The analysis begins by setting a non-zero and positive initial correlation between $p_x$ and $p_y$. The intersection of $p_e$ with the previously set positive value of correlation between inputs implicitly assumes that there is a shift in the value of correlation to arrive at the minimum MSE.
\\ \textit{a) With existing $SCC(X,Y)=1$:  }For positively correlated numbers, AND gate implements $p_z=min(p_x,p_y)$. We counter the effect of transient error on the circuit by introducing a desired $SCC_i$ obtained using following derivations. \begin{equation}\footnotesize \label{eqn:ISCC_m+1}
 I_{SCC{_m}(+1)}= 
\begin{bmatrix}
   & -(p_y-1)(p_xSCC_i-p_x+1)
\\& p_y(SCC_i- 1)(p_x-1)-SCC_i(p_x-p_y)
\\  &  p_x(SCC_i-1)(p_y-1)
\\ & p_x-p_xp_ySCC_i(SCC_i-1)
    \end{bmatrix}^\intercal
\end {equation}
The modified output $p_{zm}$ is calculated
as
\begin{equation} \footnotesize \label{eqn:pzm4}
p_{zm} = I_{SCC_{m(+1) }\times M_{and}=}p_ep_x - p_e(p_x + p_y - 1) + p_ep_y
\end{equation}
For $p_x+p_y\leq 1$, modified $MSE$ is,
\begin{equation}\footnotesize \label{eqn:MSE_and}
\begin{split}
MSE_{and}=(p_e-p_x+p_x(1-2p_e)\{SCC_i(1-p_y)+p_y\})^2
\end{split}
\end{equation}
Differentiating Eq. \ref{eqn:MSE_and} w.r.t $SCC_i$ and equate it to $0$,
\begin{equation}\label{eqn:SCC_i_7}\footnotesize
SCC_i = \frac{p_x-p_e- p_xp_y(1-2p_e)}{p_x(2p_e- 1)(p_y-1)}
\end{equation}
\begin{equation} \label{eqn:SCC_i_8} \footnotesize
SCC_i = \frac{(p_x-p_e+2p_ep_x+p_xp_y)}{p_x-2p_ep_x-p_xp_y+2p_ep_xp_y},p_x+p_y>1
\end{equation}

\par\textit{b) Any positive intermediate correlation,  $SCC(X,Y)=0.5$}: Now consider any intermediate positive correlation between the numbers, say $+0.5$. The  function implemented by AND logic at $SCC=0.5$ is $p_z =  0.5p_x(1+p_y)$ for $p_x<p_y$. In presence of transient error in the circuit, we modify input vectors as $I_{SCC_{m}}$,
\begin{equation}\label{eq25}\footnotesize \nonumber
 I_{SCC_m(+0.5)}= 
\scriptsize \begin{bmatrix}
   & (1-p_y)(0.5p_xSCC_i-p_x+1)
\\& 0.5p_y(1-p_x)-SCC_i(0.5p_x+0.5p_y-p_xp_y)
\\  &  p_x(SCC_i- 1)(p_y - 1)
\\ &  p_x.(p_y + 0.5SCC_i -0.5p_ySCC_i)
\end{bmatrix}^\intercal
\end {equation}
\begin{equation}\footnotesize
\begin{split}
p_{zm} = I_{SCC_m(+0.5)}\times M_{and}=0.5SCC_ip_x(p_y-p_e+p_ep_y)\\
+SCC_ip_x(0.5-p_e+p_ep_y)+p_xp_y(1-2p_e)+p_e
\end{split}
\label{eqn:pzm5}
\end{equation}
\begin{equation}\footnotesize
\begin{split}
 MSE_{and} =  
p_e(1-2p_xp_y)-p_x(1-p_y)\\(0.5+p_eSCC_i)+0.5p_x(1-p_y)(1-p_e)SCC_i
\end{split}
\label{eqn:MSE_and1}
\end{equation}
 Differentiating Eq. \ref{eqn:MSE_and1} w.r.t $SCC_i$ and putting it to $0$, gives
\begin{equation}\footnotesize
SCC_i=\frac{0.5p_x(1-p_y)+2p_ep_xp_y-p_e}{0.5p_x(1-p_y)(1-3p_e)}
\label{eqn:SCC_i_10}
\end{equation}
\begin{equation}\footnotesize
SCC_i=\frac{2p_ep_x-p_x-2p_e+p_xp_y+2p_ep_xp_y}{2p_x-4p_ep_x-2p_xp_y+ 4p_ep_xp_y} , \; p_x+p_y>1.
\label{eqn:SCC_i_10a}
\end{equation}
 
\par \textit{Example $6$:} 
Consider $p_x=0.3$, $p_y=0.6$ and $p_e=0.15$. 
Thus, $p_z$ and $p_{ze}$ are $0.3$ and $0.36$. From Eq. \ref{eqn:pzm5}, $p_{zm} = 0.1 SCC_i + 0.64p_e - 0.24p_eSCC_i + 0.18$. 
We invoke Eq. \ref{eqn:MSE_and1} to obtain
$MSE = 1.6\times10^{-3}(3SCC_i+ 16p_e-6p_eSCC_i-3)^2$. Thus, for $ p_e=0.15$, MSE can be reduced to zero by injecting  $SCC_i=+0.2857$.
\par With $SCC = +0.5$ between inputs,
$p_z$ and $p_{ze}$ are $0.24$ and $0.32$. Using Eq. 37, $p_{zm} = 0.1SCC_i + 0.64p_e - 0.24p_eSCC_i + 0.18$. Thus, $ MSE =\frac{(3SCC_i+32p_e-9p_eSCC_i-3)^2}{40000}$. Thus unlike the previous case,  MSE can be reduced to zero only for $p_e<0.15$ by injecting suitable positive $SCC_i$.
\vspace{2mm}
\par (ii) \textit{XOR gate:} We assume  a positive definite correlation between inputs of an XOR gate and using similar analysis we derive the condition for minimum MSE.

\vspace{1mm}
\par \textit{a) With existing $SCC(X,Y)=1$:} With positively correlated inputs, the XOR gate implements $p_z= F_{+1}=\lvert p_x - p_y \rvert$. 
The deviation  under the error scenarios can be encountered by finding a suitable operating point of the circuit by defining $SCC_i$ using the following derivations.

We modify the output by introducing the desired correlation $SCC_i$ such that,
\begin{equation}\label{eqn:pzm6}\footnotesize
\begin{split}
p_{zm}=p_e+p_x+p_y-2SCC_ip_x-2p_ep_x-2p_ep_y-2p_xp_y\\+4p_ep_xSCC_i +2p_xp_ySCC_i+4p_ep_xp_y-4p_ep_xp_ySCC_i
\end{split}
\end{equation}
\begin{equation}\footnotesize
\begin{split}
MSE_{xor}=(p_e+2p_x-2p_e(p_x+p_y)-\\2p_x(1-2p_e)\{SCC_i(1+p_y)+p_y\})^2
\end{split}\label{eqn:MSE_xor}
\end{equation}
Differentiating Eq. \ref{eqn:MSE_xor} w.r.t $SCC_i$ and equating it to $0$,
\begin{equation}\footnotesize \label{eqn:SCC_i_11}
\begin{multlined}
SCC_i= \frac{p_e(1-2p_x)(1-2p_y)+2p_x(1-p_y)}{2p_x(2p_e-1)(p_y-1)}
\end{multlined}
\end{equation}
\begin{equation}\footnotesize \label{eqn:SCC_i_12}
\begin{multlined}
SCC_i= \frac{p_e-2p_x+2p_e.p_x-2p_e.p_y+2p_x.p_y}{2p_x -4p_e.p_x-2.p_x.p_y+4p_e.p_x.p_y}
\end{multlined}
\end{equation}
\textit{Example $7$:} Consider $p_x$ = 0.3, $p_y$ = 0.6, $p_e = 0.15$. Thus,
$p_z = 0.3$ and $p_{ze} =0.36$ and $p_{zm} =  0.48p_eSCC_i-0.0800p_e-0.24SCC_i +0.54$.
Now, $MSE = \frac{(9SCC+ 2p_e - 18p_eSCC - 6)^2}{625}$, which can be reduced to zero when $SCC_i = 0.933 $.
\par  \textit{b) Any positive intermediate correlation,  $SCC(X,Y)=0.5$:}
In this case, the resultant function is $ p_z= F_{+0.5}= -p_y(p_x - 1)$ when $p_x<p_y$. The error-free output is obtained using modified output defined by input vectors in $I_{SCC_{m(+0.5)}}$.
\begin{equation}\label{eq57}\nonumber\footnotesize
\begin{split}
p_{zm} = p_e+p_x+p_y-5SCC_ip_x-2p_ep_x-2p_ep_y-2p_{xp_y}\\+10p_{ep_x}SCC_i+ 5SCC_{p_x}p_y+4p_ep_{xp_y}-10p_ep_{xp_y}SCC_i
\end{split}
\end{equation}
\begin{equation}\scriptsize \label{eqn:MSE_xor1}
\begin{split}
MSE_{xor} = 0.25(2p_e+2p_x-3p_xSCC_i-4p_ep_x-4p_ep_y-2p_xp_y\\+5p_ep_xSCC_i+ 3p_xp_ySCC_i+8p_ep_xp_y-5p_ep_xp_ySCC_i)^2
\end{split}
\end{equation}
Differentiating Eq. \ref{eqn:MSE_xor1} w.r.t $SCC_i$ and equate it to $0$,
\begin{equation}\footnotesize \label{eqn:SCC_i_13}
\begin{multlined}
SCC_i = \frac{2p_e(1-2p_y)(1-2p_x)+2p_x(1-p_y)}{p_x(5p_e-3)(p_y-1)}, p_x+p_y\leq 1
\end{multlined}
\end{equation}
\begin{equation}\footnotesize \label{SCC_i_14}
\begin{multlined}
SCC_{i}= \frac{p_e-p_x-2p_e.p_y+p_x.p_y+2p_{ep_x}p_y}{2p_x -4p_e.p_x-2p_x.p_y+4p_{p_x}.p_y}, p_x+p_y>1
\end{multlined}
\end{equation}
With positively correlated inputs the induced $SCC_i$ can generally be written in the form,
\begin{equation}\footnotesize \label{SCC_i_15}
\begin{multlined}
SCC_i =  \frac{p_e(1-2p_x)(1-2p_y)+2SCCp_x(1-p_y)}{-p_x(p_y-1)(SCC-p_e-3SCCp_e+1)}
\end{multlined}
\end{equation}
\textit{Example $8$:} Let $p_x= 0.3$, $p_y= 0.6$ and $p_e = 0.15$. Thus,
$p_z = 0.42 $ and $p_{ze} = 0.444 $. And $p_{zm} = 1.12p_eSCC - 0.08p_e - 0.56SCC + 0.54 $. Thus, ${MSE = 0.0016(14SCC_i+2p_e-28 p_e SCC_i-3)^2}$,
which can be reduced to zero when $SCC_i =+ 0.275 $.
\vspace{2mm}
\par
\begin{figure}
     \centering
     \begin{subfigure}[b]{0.4\textwidth}
         \centering
         \includegraphics[width=\textwidth]{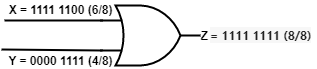}
         \caption{ $F_{-1} = min(p_x+p_y,1)$}
         \label{}and
     \end{subfigure}
     \hfill
     \begin{subfigure}[b]{0.4\textwidth}
         \centering
         \includegraphics[width=\textwidth]{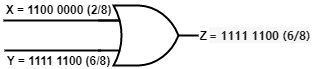}
         \caption{$F_{+1} = max(p_x,p_y)$}
         \label{}
     \end{subfigure}
     \caption{OR gate implementing different functions when inputs have different correlation status.}
     \label{fig:ORgate_correlation}
\end{figure}

 (iii) \textit{OR gate:} Consider an OR gate with positively correlated inputs. With different correlation status OR gate implements different stochastic functions as shown in Fig. \ref{fig:ORgate_correlation}. Using \textit{ReCo} analysis we try to derive the condition for achieving minimum MSE when the gate is assumed to be inflicted with transient noise.
\vspace{2mm}

\vspace{1mm}
a) \textit{With existing \textbf{$SCC(X,Y)=1$}:} The OR gate implements $p_z= max(p_x,p_y)$ when two numbers are positively correlated. We can similarly find the operating $SCC_i$  to obtain minimum $MSE$ under error scenarios.
\begin{equation}\label{eqn:pzm7} \footnotesize
\begin{split}
\therefore p_{zm} = p_e+p_y-p_ep_x-2p_ep_y+p_ep_xSCC_i+p_ep_xp_y\\-p_ep_xp_ySCC_i- p_x(SCC_i-1)(p_e-1)(p_y-1)
\end{split}
\end{equation}
\begin{equation} \label{eqn:MSE_or1} \footnotesize
\begin{split}
MSE_{or} =(p_e+p_x-p_xSCC_i-2p_ep_x-2p_ep_y-p_xp_y\\+2p_ep_xSCC_i+p_xp_ySCC_i+2p_ep_xp_y-2p_ep_xp_ySCC_i)^2
\end{split}
\end{equation}
\begin{equation}\footnotesize \nonumber 
SCC_i = \frac{p_e+p_x-2p_ep_x-2p_ep_y-p_xp_y+ 2p_ep_xp_y}{p_x(2p_e-1)(p_y-1)}
\end{equation}
\begin{equation} \footnotesize  \label{eqn:SCC_i_17}
\therefore SCC_i = \frac{p_e-p_x-2p_e.p_y+p_x.p_y}{p_x-2p_e.p_x- p_x.p_y+2p_e.p_x.p_y}
\end{equation}
\textit{Example $9$:} Consider $p_x$ = 0.3 and $p_y$ = 0.6,
$p_z = 0.6$ and $p_{ze} =0.57 $ for $p_e = 0.15$. $p_{zm}$ = $0.24p_eSCC-0.44p_e-0.12SCC+ 0.72$
with the help of above equations (70) and (71),
$MSE = \frac{(9SCC + 22p_e - 18p_eSCC- 6)^2}{40000}$. Thus,
MSE can be reduced to zero for $p_e=0.15$ when $SCC_i = +0.432$.
\vspace{2mm}
\par b) \textit{Any positive intermediate correlation,  $SCC(X,Y)=+0.5$:} In this case, OR gate implements $p_z= 0.5p_x + p_y - 0.5p_xp_y$. To find the operating $SCC_i$  to minimize $MSE$ under given error scenarios we take help of following derivations.
\begin{equation}\footnotesize \label{eqn:pzm8}
\begin{split}
p_{zm}=p_e+p_y-p_ep_x-2p_ep_y+ 0.5p_ep_xSCC_i+p_ep_xp_y\\\- 0.5p_ep_xp_ySCC_i-p_x(SCC_i-1)(p_e- 1)(p_y-1)
\end{split}
\end{equation}
Using $p_{zm}$ from the previous case,
\begin{equation} \footnotesize \label{eqn:MSE_or2}
\begin{multlined}
MSE_{or}=0.25(2p_e+p_x-2p_xSCC_i-4p_ep_x-4p_ep_y-p_xp_y\\+ 3p_ep_xSCC_i+2p_xp_ySCC_i+4p_ep_xp_y-3p_ep_xp_ySCC_i)^2
\end{multlined}\end{equation}
\begin{equation}\footnotesize \label{eqn:SCC_i_18}
\begin{multlined}
SCC_i = \frac{2p_e+p_x- 4p_ep_x-4p_ep_y-p_xp_y+4p_ep_xp_y}{(2p_x-3p_ep_x- 2p_xp_y +3p_ep_xp_y)}
\end{multlined}
\end{equation}
Similarly, for $p_x+p_y>1$
\begin{equation}\footnotesize \label{eqn:SCC_i_19}
\begin{multlined}
SCC_i =\frac{2p_e-p_x-2p_ep_x-4p_ep_y+p_xp_y+ 2p_ep_xp_y}{2p_x-4p_ep_x-2p_xp_y+4p_ep_xp_y}
\end{multlined}
\end{equation}
\textit{Example $10$:} Consider $p_x$ = 0.3, $p_y$ = 0.6 and $p_e = 0.15$. Thus, $p_z = 0.66 $ and $p_{ze} = 0.612$. $p_{zm}$ = $0.24p_eSCC-0.44p_e-0.12SCC+ 0.72$. Using Eq. \ref{eqn:MSE_or2}, $MSE = \frac{(9SCC + 22p_e - 18p_eSCC - 3)^2}{40000}$. Thus, MSE can be reduced to zero for above  $p_e$ when $SCC_i = -0.05 $.
\par Thus for an OR gate, induced correlation are mostly obtained in the positive range for the extreme case of initial $SCC = +1$ as shown in Fig. \ref{fig:MSE_OR gate}(c), while for initial $SCC = +0.5$, the injected $SCC_i$ values are mostly obtained in the negative range except for $p_e\leq 0.1$. The simulation results for initial $SCC=0.5$ are shown in Fig. \ref{fig:MSE_OR gate}(d).
For any positive correlation $SCC_1$ the expressions can be generalized as:
\begin{equation}\footnotesize \label{SCC_i_16}
SCC_{i}=\frac{p_e+SCCp_x(1-p_y)-2p_e(p_x+p_y)+ 2p_ep_xp_y}{p_x(1-p_y)(1-p_e)-SCC_1.p_e.p_x(1-p_y)}
 \end{equation}
\subsection{The proposed error detector circuit}
\begin{figure}
    \centering
    \includegraphics[width=0.45\textwidth]{ 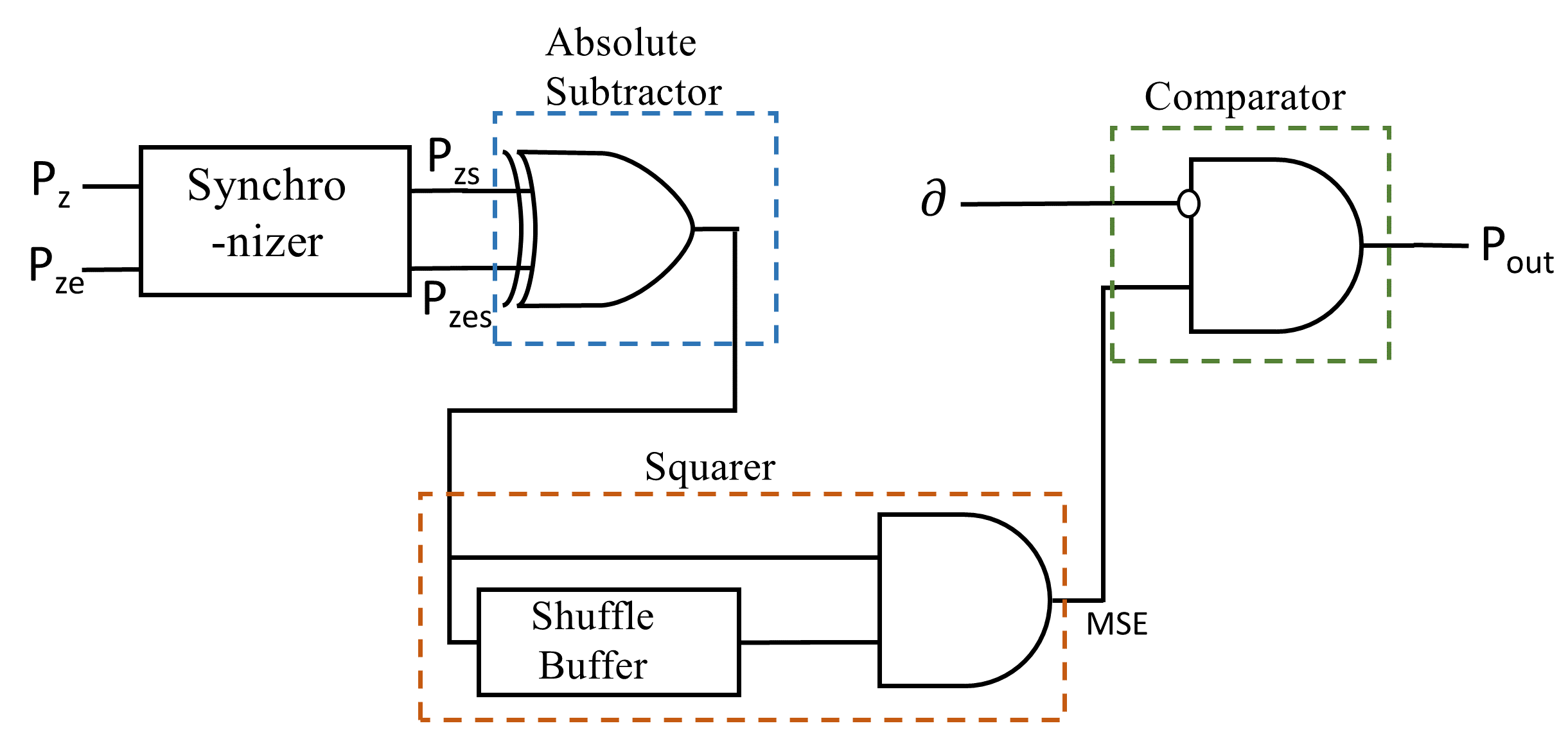}
    \caption{  The Error Detector Circuit.}
    \label{fig:ReCo_AND gate}
\end{figure}
\par
\begin{figure}
    \centering
    \includegraphics[width=0.35\textwidth]{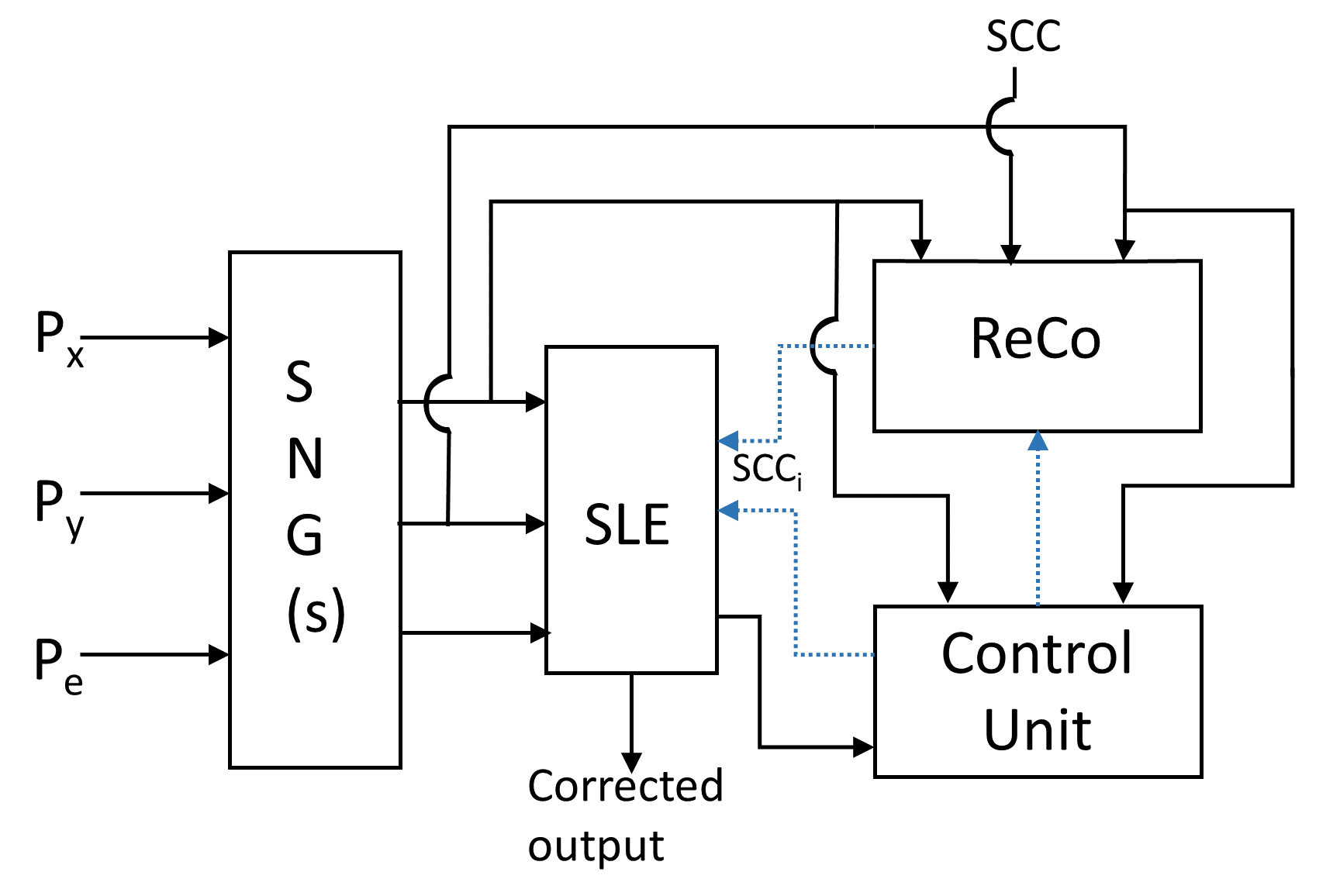} 
    \caption{System-Level representation of Error Correction mechanism.}
    \label{fig:SystemLevelRepresentation}
\end{figure}

 Consider an SLE in a noisy environment. The system level representation of the error correction mechanism for such an SLE is shown in Fig. \ref{fig:SystemLevelRepresentation}. The inputs to the unit are $p_x$ and $p_y$,  error $p_e$ and output is the desired value $p_{zm}$. The control circuit or the error detector circuit is used to determine the amount of deviation of an erroneous output from an error-free output. It comprises a subtractor, a squarer and a comparator which determines the amount of error that is to be reduced. Auxiliary circuits like the shuffle buffer and the synchronizer are used to adjust the correlation between the input bitstreams. The output of the control unit  is fed to the \textit{ReCo} block consisting of a correlator circuit that generates the desired $SCC_i$ to  minimize MSE. The control circuit is described below.

\subsubsection{Synchronizer}
\label {synchronize}
The synchronizer \cite{lee2018correlation} is a finite state machine that pairs up a maximum number of input bits to $00$ or $11$, restoring their respective probabilities. This  unit is placed between two uncorrelated sequences, $p_{ze}$ and $p_z$ to introduce positive correlation between sequences $p_{{ze}_{s}}$ and $p_{z_{s}}$.  If the bits in $p_{ze}$ and $p_{z}$ are equal, then the corresponding bits are given as output. When bits are dissimilar, depending upon input values $1$ or $0$, if they are in state $S1$, are changed to $S0$ and $S2$, both $0$ or both $1$ are given as output.  In this process, the probabilities of $p_{ze}$ and $p_{z}$ are kept  unchanged.
\subsubsection{Subtractor}
 The difference between the error value $p_{ze}$ and the actual value $p_{z}$ is calculated to check the amount of deviation. The XOR gate performs absolute subtraction i.e, $p_{diff}=|p_{z_s}-p_{ze_s}|$, when $p_{z_s}$ and $p_{ze_s}$ are positively correlated \cite{alaghi2013exploiting} which is done using a synchronizer.  
\subsubsection{Squarer}
\label{sqr}
Multiplication of two uncorrelated SNs is performed by an AND gate, but fails to implement the squaring operation \cite{alaghi2013exploiting} when the same input sequence is given. When $p_{diff}$ is squared to obtain the MSE, it is necessary to minimize the correlation between SNs. A shuffle buffer circuit is used \cite{lee2018correlation} to reduce the correlation between inputs to obtain the accurate squaring operation. It includes a multiplexer, $3$ D Flip-flops and a Random Number Generator to generate numbers between $0$ and $1$ \cite{lee2018correlation}.
\par
\subsubsection{Comparator} The stochastic comparator shown in Fig. \ref{fig:ReCo_AND gate} compares the obtained MSE with $\delta$ ($0.0001$). It is  based on the fact \cite{alaghi2013exploiting} that when two correlated inputs are given to an AND gate with an inverter to one of its inputs the stochastic function implemented  is given by $p_{out}=max((MSE-\delta),0)$.  Thus, when $MSE$ which is representative of the error in computation is lesser than the error-tolerance of the circuit $\delta$, a bitstream of $0$'s is obtained at the output of the comparator which indicates that the output is obtained satisfactorily.
\section{Formalization of Reco Technique for stochastic circuits}
The next step is to formally apply the proposed technique in a combinational circuit. 
\begin{figure}[htp]
       \centering
       \includegraphics[height=2.8 cm]{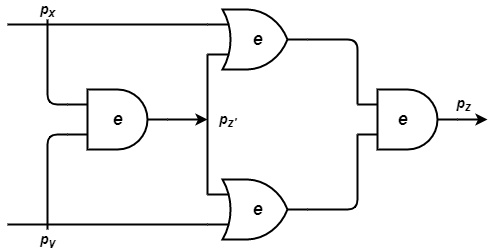}
        \caption{A 2 input multi-level circuit.}
       \label{fig:Two_input_multilevel}
   \end{figure}
 For a two-input multilevel circuit as shown in Fig.\ref{fig:Two_input_multilevel}, we use circuit PTM  obtained as,
$M_{ckt} = (F_2 \otimes F_2).(M_{and} \otimes I \otimes I).(F_2 \otimes I \otimes I).(M_{or} \otimes M_{or}).M_{and}. $
\par Consider $p_x = 0.3$ and $p_y = 0.6$. The error free output is $p_z$ = 0.29.
If $p_e=0.125$, the observed output, $p_{ze} = 0.33.$
Using the proposed method,
$MSE = (0.066SCC + 0.0388)^2$ which shows that  MSE can be reduced to zero for $SCC \approx -0.58$.
\par We will now discuss two distinct cases of fault correction in multi-input multi-level circuits.
\subsection{ReCo I:Error minimization  for Multi-input Single output (MISO) SLCs} 
\par
\begin{figure}[htp]
       \centering
       \includegraphics[height=5.2cm]{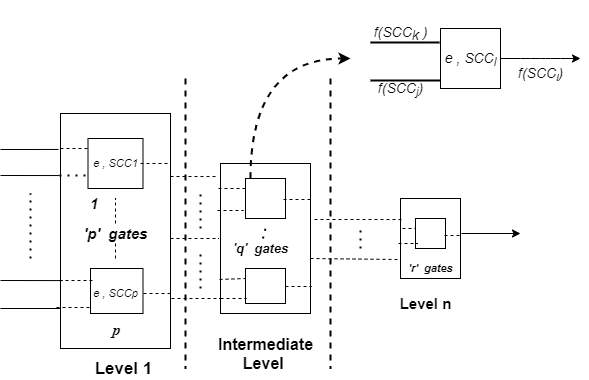}
       \caption{ Multi-level block representation  of MISO SLC.}
       \label{fig:Intermediate_gate_in_the_circuit}
\end{figure}

 Let us consider  correlation-sensitive blocks interconnected in a fashion as shown in Fig. \ref{fig:Intermediate_gate_in_the_circuit}. The quantification of SCC is only available for two signals in literature, so we consider two input gates in the circuit model. The block diagram consists of $i=1,2,..,n$ levels and each level consists of multiple two input gates. Consider one or more gates at different levels are subject to transient faults which result in an increased MSE at the output. The MSE is minimized by selecting suitable candidates for \textit{ReCo} analysis using the proposed Algorithm 2 which is discussed below.
 \vspace{0.5mm}
\par \textit{Definition 4: An observed error at the primary output(s) driven by one or more gates, can be minimized by injecting correlation at prioritized input gates defined by the faulty gates in the error path P.}
\vspace{0.5mm}
\par Assume correlation between input pairs for gates at level $1$ as $SCC_{11}$,$SCC_{12}$,$SCC_{13}$,..,$SCC_{1p}$, where $p$ is the number of two input gates at level 1. We can represent the output of each gate from  level $1$ in terms of $SCC$ to become functions of corresponding $SCC$. Thus, $p_{z_{11}}$=$f(SCC_{11})$, $p_{z_{12}}$=$f(SCC_{12})$, $p_{z_{13}}$=$f(SCC_{13})$,..., $p_{z_{1p}}$=$f(SCC_{1p})$. 
These outputs are again inputs to certain gates in the next level of the circuit. The output from any gate in the intermediate level, is thus, in turn  a function of $SCC$ of primary input pairs. \par Let $i^{th}$  intermediate level  consists of $q$ interconnected gates. The output $Z_{i1}$ from gate 1 in $i^{th}$ level is $p_{Z_{i1}}=f(p_{Z_{(i-1)k}},p_{Z_{(i-1)j}})$ where $Z_{(i-1)k}$ and $Z_{(i-1)j}$ are the outputs from $k^{th}$ and $j^{th}$ gate at the $(i-1)^{th}$ level. Thus, $p_{Z_{i1}}$=$f(p_{Z_{(i-2)l}},p_{Z_{(i-2)m}}, p_{Z_{(i-2)q}}, p_{Z_{(i-2)r}})$, where $p_{Z_{(i-2)l}}$ and $p_{Z_{(i-2)m}}$ are the outputs of $l^{th}$ and $m^{th}$ gate of $(i-2)^{th}$ level that are inputs to $k^{th}$ gate in $(i-1)^{th}$ level and $p_{Z_{(i-2)q}}$ and $p_{Z_{(i-2)r}}$ are the outputs from $q^{th}$ and $r^{th}$ gate of $(i-2)^{th}$ level that are connected to $j^{th}$ gate in $(i-1)^{th}$ level. These outputs are again functions of SCCs of their corresponding inputs, such that $p_{{Z}_{i1}}$=$f(SCC_{(i-2)l},SCC_{(i-2)m},SCC_{(i-2)q},SCC_{(i-2)r})$. Thus, tracing the interconnected sub-networks, $p_{{Z}_{i1}}$ is written as, $p_{{Z}_{i1}}=f(SCC_{11}$,$SCC_{12}$,$SCC_{13}$,..,$SCC_{1p})$ of traced inputs at the primary level.
\par Let transient faults at a certain level contribute to shift in the desired $SCC$ at the input assumption of succeeding levels leading to erroneous results.
We take into account the intermediate correlation $SCC_{i-1}$ present between  $Z_{(i-1)k}$ and $Z_{(i-1)j}$ if the circuit is non-faulty. Let, $Z'_{(i-1)k}$ and $Z'_{(i-1)j}$ are the modified values on account of errors from $k^{th}$ and $j^{th}$ faulty gates of level $(i-1)$.  These result in a change in the number of 1's present in the bitstream. Let, probability of $Z_{(i-1)k}$  is changed from $n_{(i-1)k}$ to $n'_{(i-1)k}$ and $Z_{(i-1)j}$ is changed from $n_{(i-1)j}$ to $n'_{(i-1)j}$. Eventually, $SCC_{i1}$ is modified to $SCC'_{i1}$ which modifies the probablity of $Z_{i1}$ from $p_{z_{(i-1)j}}$ to $p_{z'_{(i-1)j}}$. Thus, $p_{z'_{(i-1)j}}= (1+SCC'_{(i-1)j})p_{z'0_{(i-1)j}}-SCC'_{(i-1)j}p_{z'(+1)_{(i-1)j}} =n'_{(i-1)j}/n $.
Similarly,  $p_{z'_{(i-1)k}}$ can be written as
 $p_{z'_{(i-1)k}}= (1+SCC'_{(i-1)k})p_{z'0_{(i-1)k}}-SCC'_{(i-1)k}p_{z'(+1)_{(i-1)k}} =n'_{(i-1)k}/n $.

Thus the effect of transient error is reflected in an overall change in the  probability of SNs. This suggests the dependence of MSE on SCC. This instigated us to find a suitable technique that adapts to the change in the assumption of correlation and try to minimize the observed error at the output by using \textit{ ReCo}. It relocates $SCC$ at several levels to counterbalance the change in initial $SCC$ due to transient faults. We trace faults and determine gates subjected to faults and remodel SCCs by modifying input vectors of the primary inputs of the sub network in level 1.
\begin{figure}[htp]
       \centering
       \includegraphics[height=3.44cm]{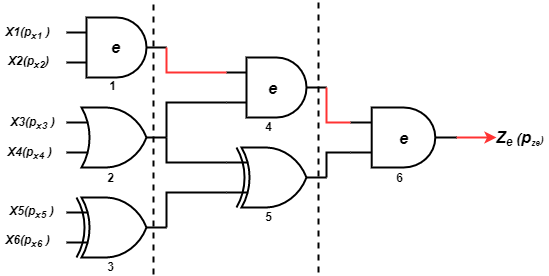}
       \caption{A sample MISO circuit showing gates $G1,G4,G6$ faulty.}
       \label{fig:Intermediate_gate_miso}
\end{figure}
\par  In noisy operating conditions, error is assumed to be primarily contributed by one or more faulty gates in the circuit. For a multilevel MISO circuit shown in Fig. \ref{fig:Intermediate_gate_miso}, the number of SLEs that undergo  \textit{ReCo} correction primarily depend on the number of faulty gates and the probability of error. Thus, it is important to identify faulty paths in the circuit.  We generalize the procedure for fault correction in MISO circuits as follows: 
\begin{enumerate}
    \item Check if the  MSE of the circuit is within the tolerable limit $\delta (<=10^{-3})$ or not. If not,
\item \textit{\textbf{Determine Faulty gates}}: Generate \textit{T} input test vectors \textit{n} times and the output at each gate is observed. The error rate is estimated by the number  of faulty outputs for \textit{n} inputs using \textit{\textbf{FaultEvaluation}}($CIRCUIT$). 
\item \textit{\textbf{Determine input SLEs corresponding to faulty output node:}} Input gates  that are connected to the faulty output node, denoted by \textit{\textbf{IsConnected}}(), are selected and are stored in an array $FC\_I\_gate$.
\item \textit{\textbf{Register  SLEs based on priority:}}  The gates  based on their priority values from left (highest) to right (lowest); \{$XOR,AND,OR$\} are sorted using $\textit{\textbf{PrioritySort}}()$ are then stored in an array $S\_FC\_I\_gate$.
\item \textit{\textbf{Selection of SLEs for \textit{     ReCo}:}} Pop gate from the priority list and alter $SCC$ using  $\textit{\textbf{ReCo}}()$ (Algorithm $1$). Calculate if, $ MSE\leq\delta$, return corresponding $SCC_i$. If not, dual combination of  SLEs.
 \item \textit{\textbf{Combine SLEs: }} Store MSE in  order of increasing magnitude in the sorted list  [\ $L\_gate, L\_MSE$]\ using $\textit{\textbf{Sort}}()$.  Club next SLEs from the sorted list till the desired MSE is reached.
  \end{enumerate}
 \par From the previous discussion, it can be inferred that the XOR gate is most susceptible to changes in correlation and can reduce the overall MSE substantially compared to other correlation sensitive SLEs. So XOR gate line up highest in the correlation-sensitivity index. If a single gate does not suffice then we proceed for combinations from the list \textbf{\textit{L\_MSE[]}}. Suppose, $\textit{\textbf{PrioritySort}}()$ list consists of \{$XOR, AND1, AND2$\} and $AND2$ generates less MSE compared to $AND1$, then we combine $XOR$ and $AND2$ to find minimum MSE. This condition is often guided by the position of the gate(s) in the circuit. The whole analysis is carried out to improve the accuracy and reduce the number of correlators in the circuit.

\begin{figure*}
      
     \centering
     
     \begin{subfigure}[b]{0.192\textwidth}
         \centering
         \includegraphics[width=\textwidth]{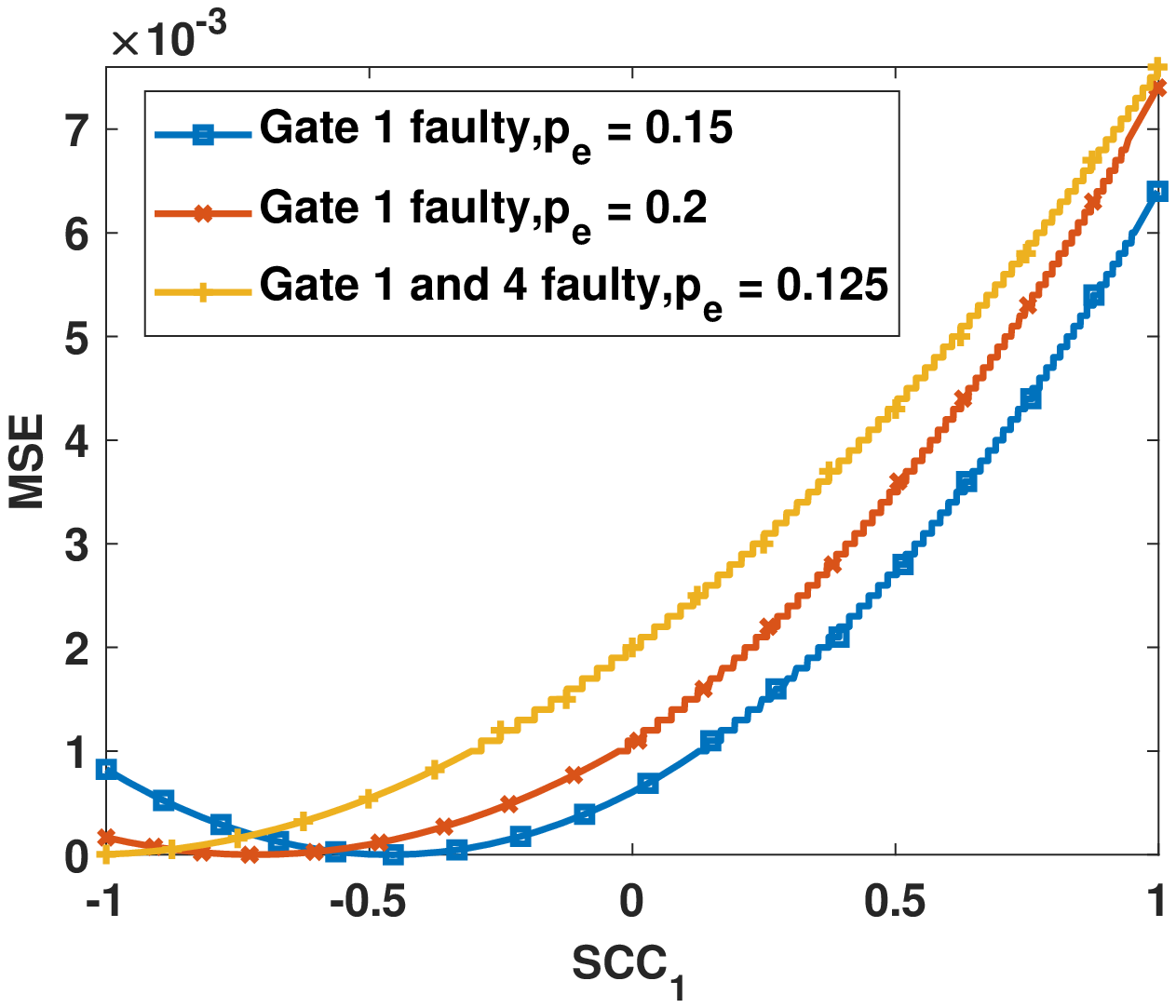}
         \caption{}
     \end{subfigure}
     \hfill
     \begin{subfigure}[b]{0.192\textwidth}
         \centering
         \includegraphics[width=\textwidth]{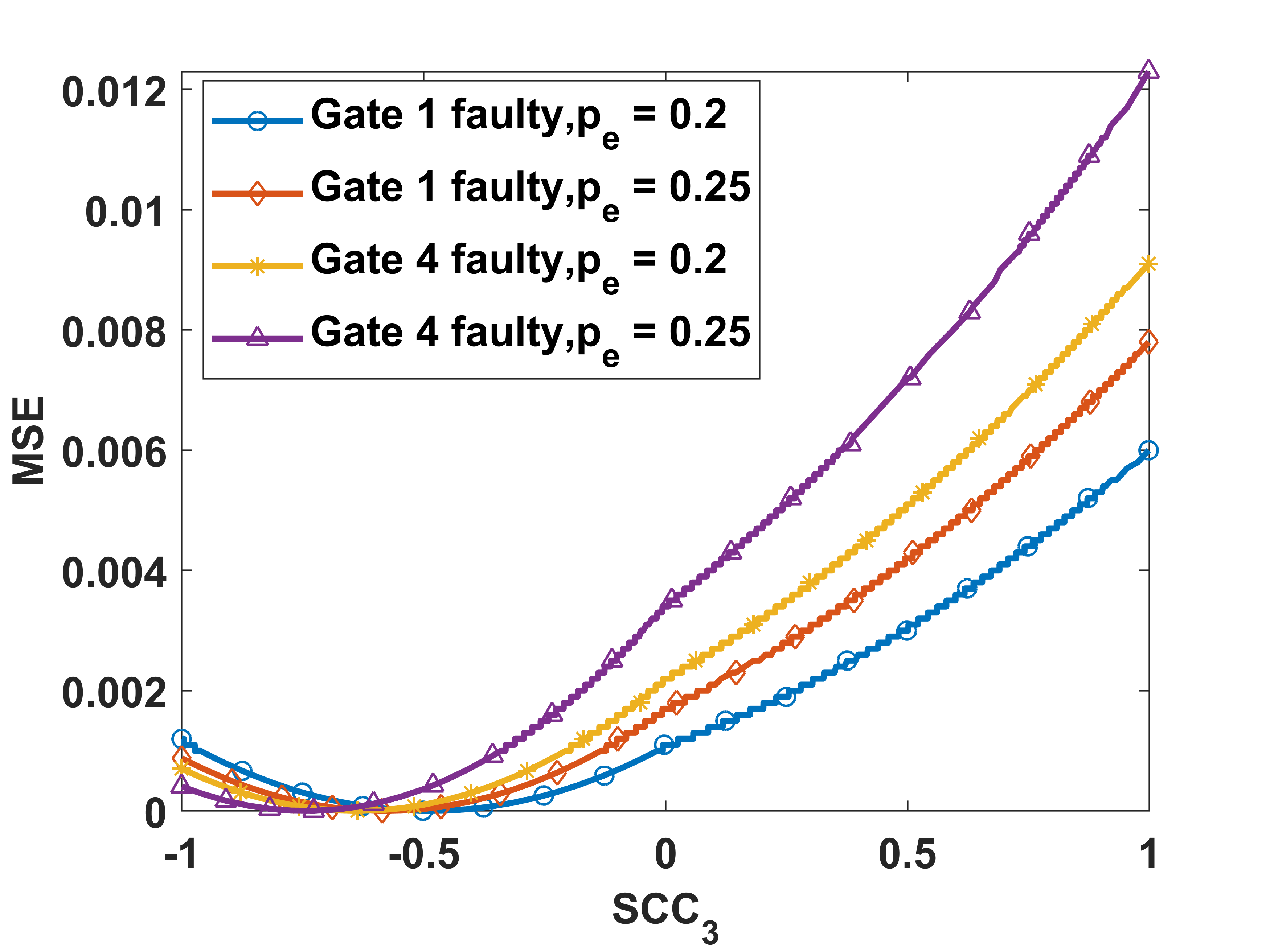}
        \caption{}
     \end{subfigure}
     \hfill
      \begin{subfigure}[b]{0.192\textwidth}
         \centering
         \includegraphics[width=\textwidth]{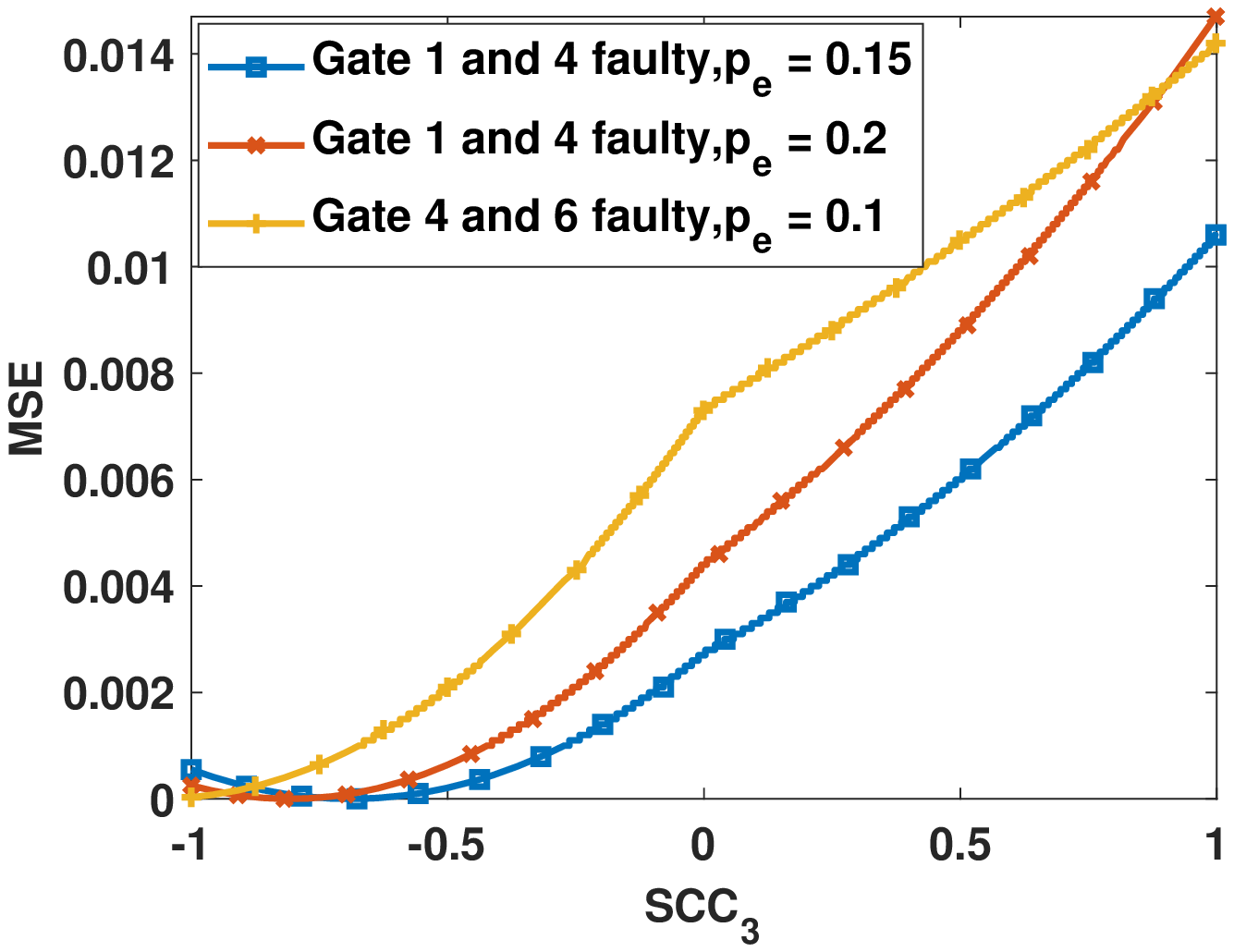}
         \caption{}
     \end{subfigure}
   \hfill
      \begin{subfigure}[b]{0.19\textwidth}
         \centering
         \includegraphics[width=\textwidth]{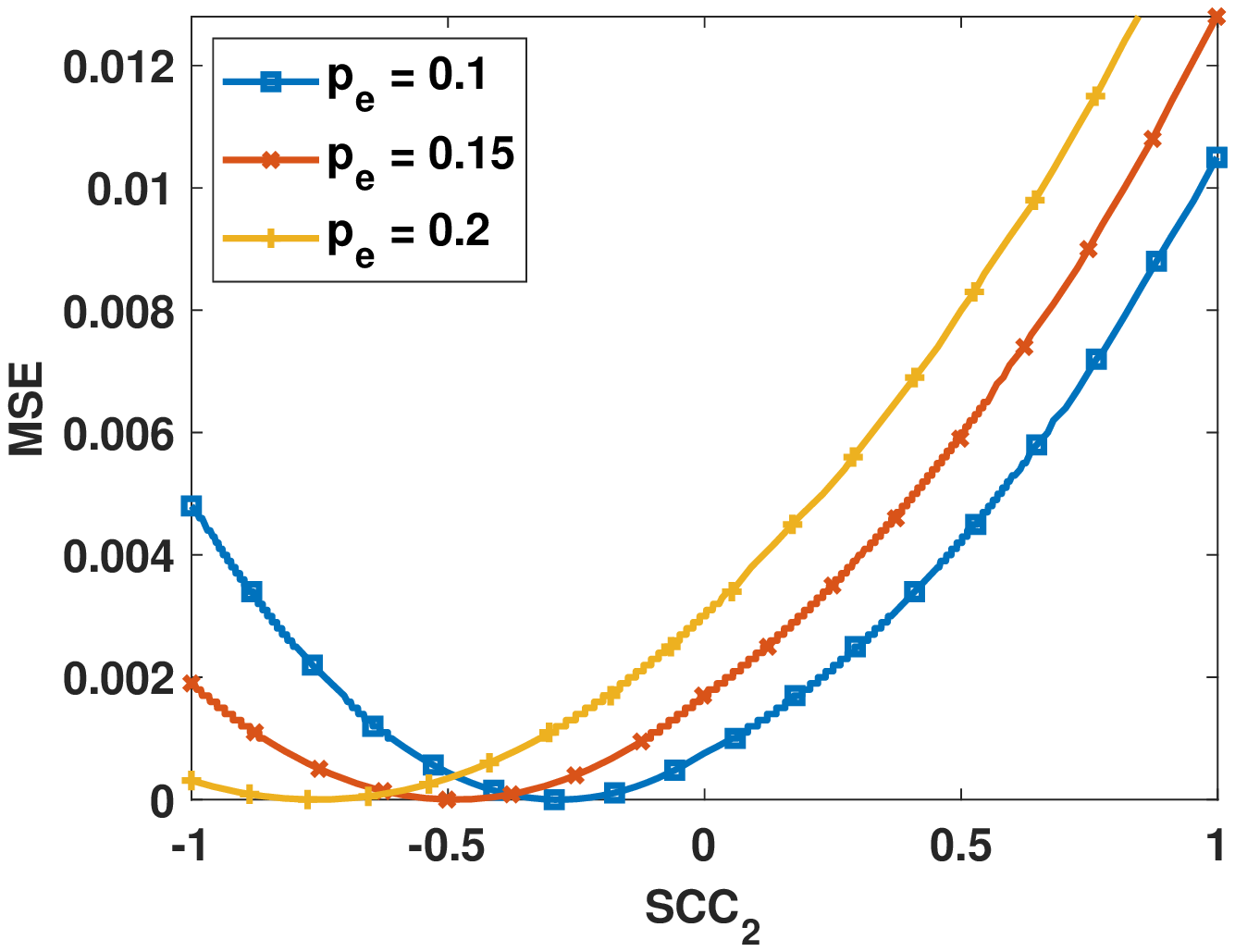}
         \caption{}
     \end{subfigure}
     \hfill
      \begin{subfigure}[b]{0.19\textwidth}
         \centering
         \includegraphics[width=\textwidth]{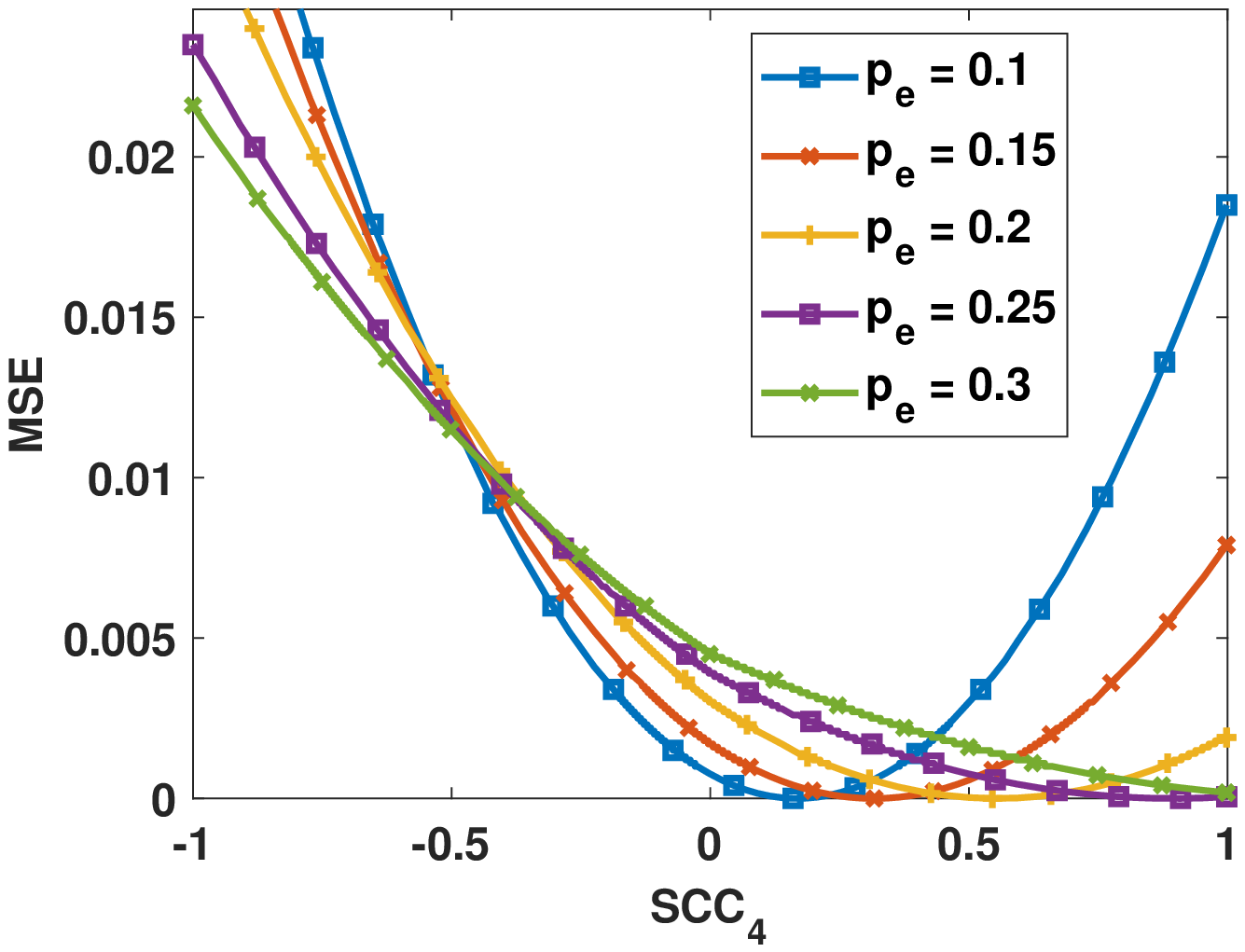}
         \caption{}
     \end{subfigure}
    \caption{Graphs showing different values of induced correlation to achieve minimum MSE; (a)-(c) MISO Circuit; (d),(e) MIMO circuit. }
    \label{fig:Reco_Varying_single_SCC}
\end{figure*}
\par It is observed that the  MSE is proportional to the number of faulty gates and the error rate $p_e$. For any error observed at the output, the error can be due to transient error at the gate itself or due to the error being propagated from the previous stage or the both. We consider different cases of fault propagation in Fig. \ref{fig:Intermediate_gate_miso} where the path P is indicated in red color.
\begin{figure}[htp]
       \centering
       \includegraphics[height=3.4cm]{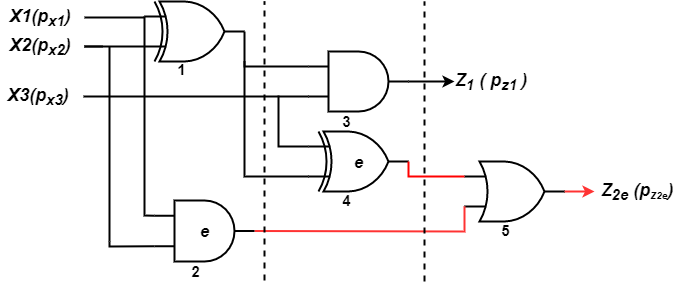}
       \caption{A sample MIMO circuit showing gates $G2,G4$ faulty.}
       \label{fig:Intermediate_gate_MIMO_circuit}
\end{figure}


\par It is observed from Fig. \ref{fig:Reco_Varying_single_SCC}(a),(b) that for a single faulty gate,  $G1$/$G4$ with $ p_e \leq 0.20$, single gate \textit{ReCo}, $SCC_1$ or $SCC_3$ will suffice to minimize MSE.  For multiple $G1,G4$ faulty  with $p_e\leq0.125$ same logic can comply. Thus, in both cases, the error can be minimized without being thoroughly guided by the priority-rule of SLE (excluding OR). But $G1,G4$ faulty at $p_e=0.2$ fault tolerance can be achieved  using prioritized SLE ($SCC_3$) only as shown in Fig.\ref{fig:Reco_Varying_single_SCC}(c) (red).
 \par For a larger number of gates faulty even at low error rates, treatment of prioritized SLE is obligatory. Thus, $G1,G4,G6$ faulty at $p_e\leq 0.125$, $MSE$ can be minimized using single gate \textit{ReCo} ($SCC_3$). But the same with $p_e=0.20$ we invoke dual \textit{ReCo} ($SCC_3, SCC_1$) as shown in Fig. \ref{fig:3D_curves_SCC_1_and_SCC_3}(d). 
 When $G4$ and $G6$  are faulty with $ p_e\leq0.25$ we invoke dual \textit{ReCo}($SCC_3,SCC_1$)  as shown in Fig. \ref{fig:3D_curves_SCC_1_and_SCC_3}(a)-(c). It is observed that there is a finite error of $0.01$ when  $G1,G4,G6$  are faulty at a rate of $ 0.25$ even after dual \textit{ReCo} of prioritized SLEs.
Table \ref{tab:Table1} is given to comprehend the nature of the analysis and the results obtained in the  proposed work.
\par It is observed that 
the possibility of error reduction is motivated by several factors; the  nature  and number of faulty gates as well as the arrangement of gates in the circuit. As AND gate is more sensitive to soft errors than the XOR gate, an AND gate in place of the XOR gate  would contribute to larger MSE. These factors coupled  with the probability of  error  play a pivotal role in determining the circuits' resilience towards soft errors. There is also a slight dependence on input probability values i.e $p_x$ and $p_y$ as indicated in Fig. \ref{fig:MSE_Varying_input_probabilities} (a)-(c).
\vspace{2mm}
\subsection{ ReCo II:: Error minimization  for Multi-input and Multi-output (MIMO) SLC}
 We have argued that \textit{ ReCo} analysis  of the prioritized SLEs can bring down MSE close to 0. We will see that in particular situations  that can deviate from this initial assumption. When non-faulty gates converge to a different output node, \textit{ReCo} analysis of primary SLEs  may give undesired results. The condition can be best described with the help of a Multi-input-Multi-Output circuit shown in Fig. \ref{fig:Intermediate_gate_MIMO_circuit}. The circuit consists of two distinct outputs $Z_1$ and $Z_2$ with probabilities $p_{z_1}$ and $p_{z_2}$. Consider two faulty gates 2 and 4 that converge to $p_{z_2}$ i.e., the output of gate 5 and the output is modified to  $p_{ze_2}$.  We assume that there are no faulty gates in the path that converges to $p_{z_1}$. One way of suppressing the propagation of faulty results to the non-faulty output node is to perform \textit{ReCo} analysis at the inputs of faulty gates only. This avoids analysis of primary SLEs that are explicitly connected to non-faulty outputs.
  \par
\begin{figure*}
     \centering
     \begin{subfigure}[b]{0.195\textwidth}
        \centering
         \includegraphics[width=\textwidth]{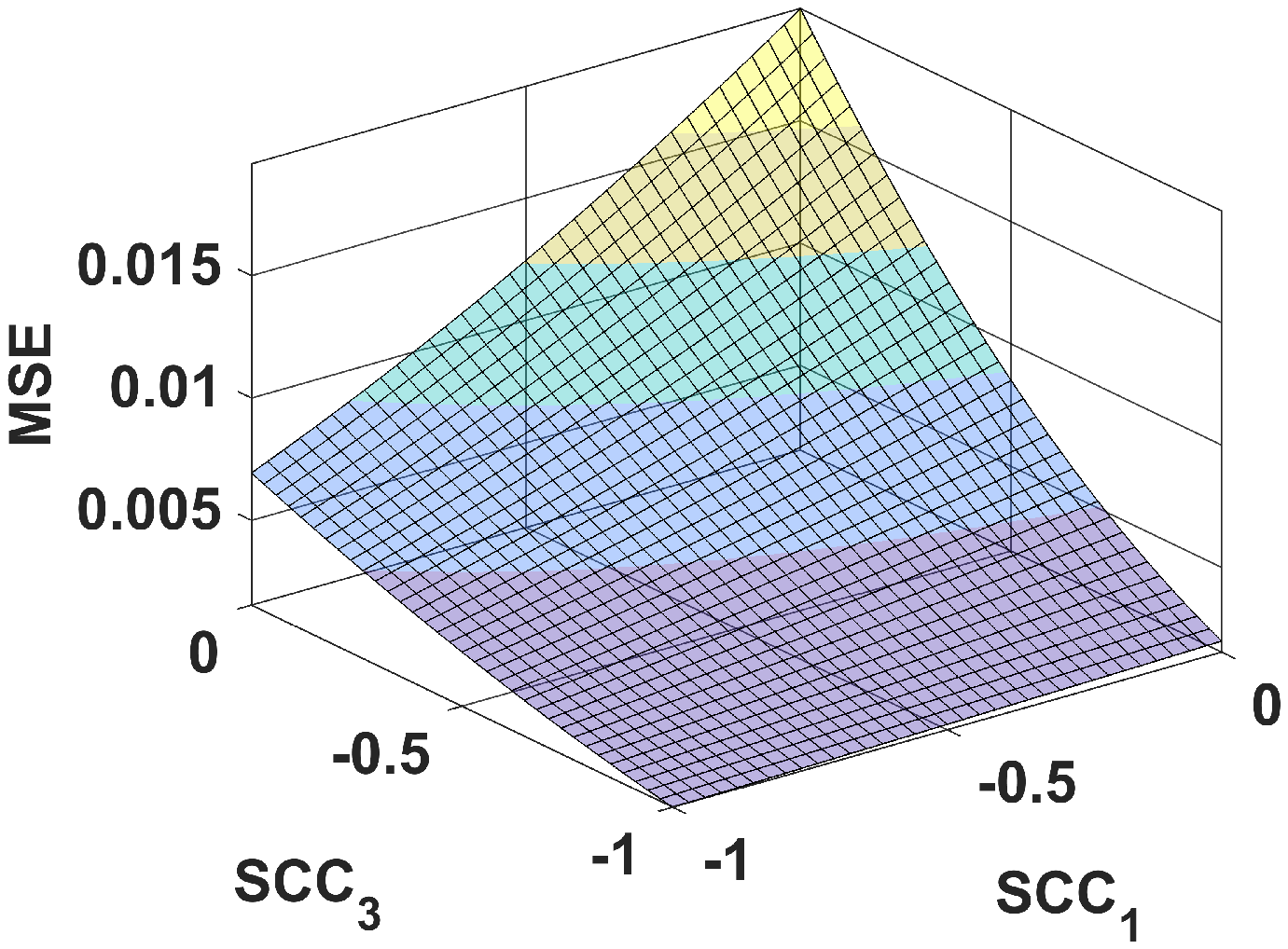}
         \caption{$p_e=0.15$}
         \label{Gate 1 and 4 faulty with p_e=0.2 varying SCC_1 and SCC_3}
     \end{subfigure}
     \hfill
     \begin{subfigure}[b]{0.195\textwidth}
         \centering
         \includegraphics[width=\textwidth]{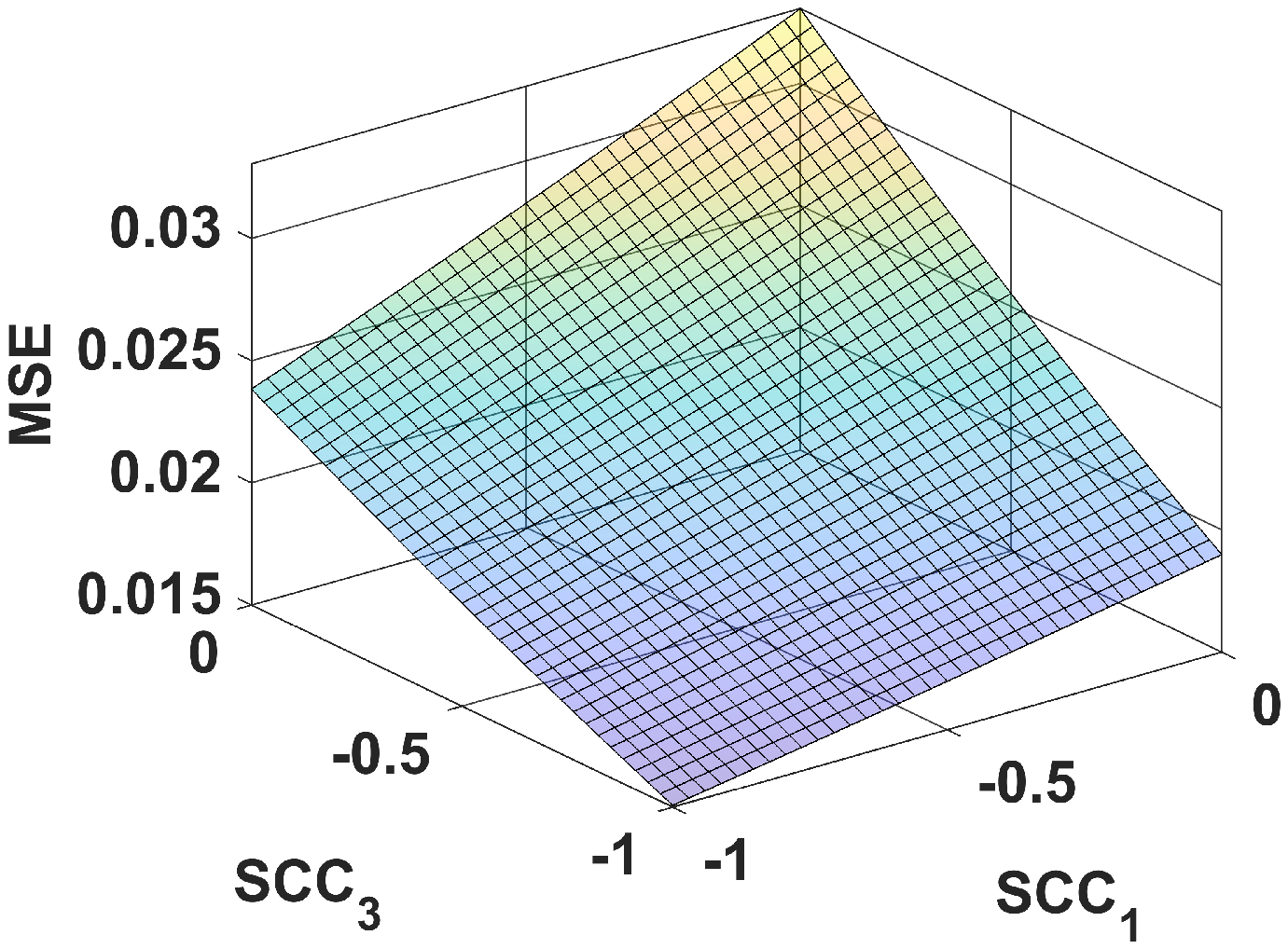}
         \caption{ $p_e=0.20$}
         \label{Gate 1,4 and 6 faulty with p_e=0.2 varying SCC_1 and SCC_3}
    \end{subfigure}
     \hfill
      \begin{subfigure}[b]{0.195\textwidth}
         \centering
         \includegraphics[width=\textwidth]{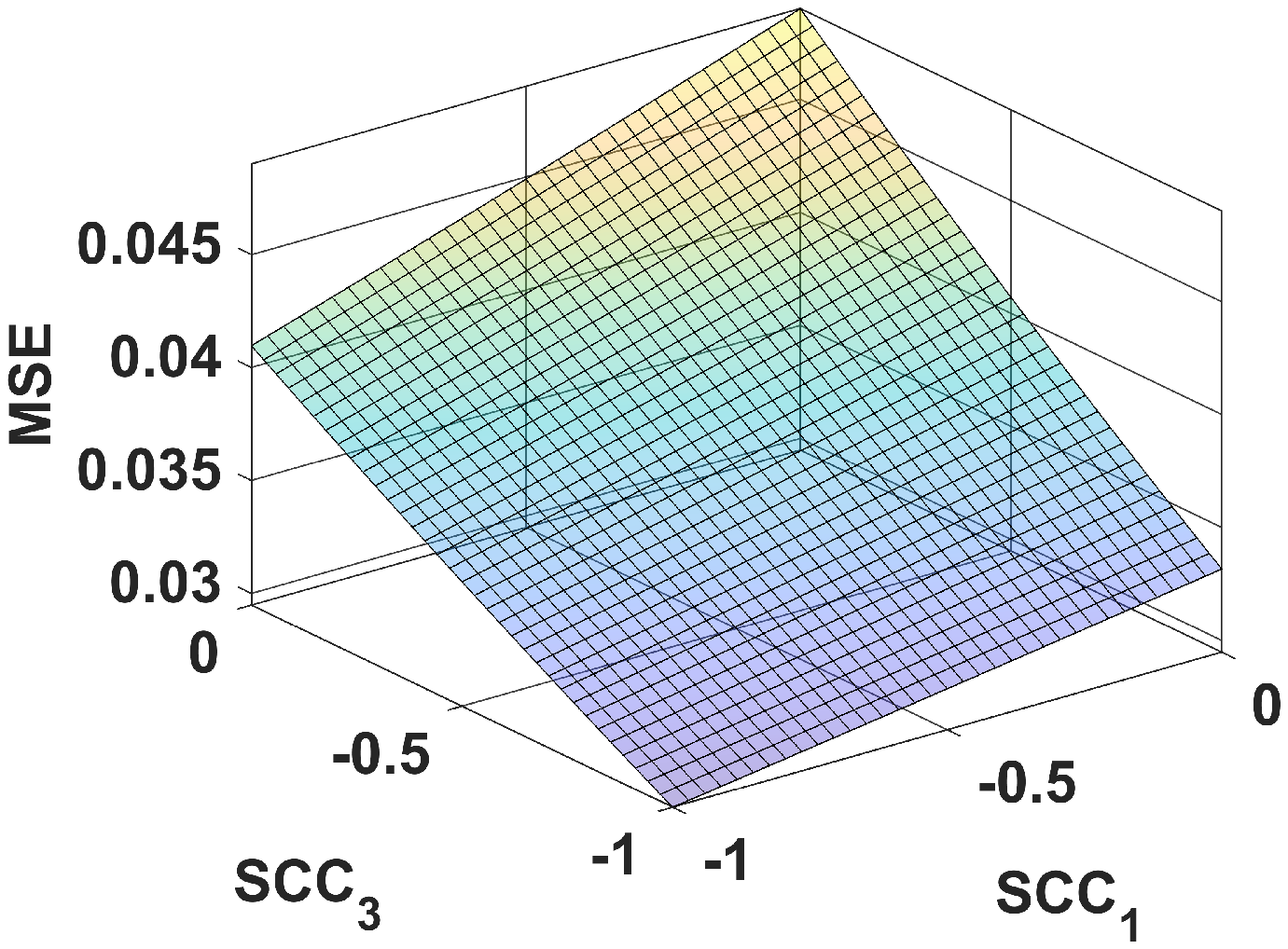}
         \caption{$p_e =0.25$}
         \label{Gate 1 and 4 faulty with p_e =0.25 varying SCC_1 and SCC_3}
     \end{subfigure}
     \hfill
    \begin{subfigure}[b]{0.195\textwidth}
         \centering
         \includegraphics[width=\textwidth]{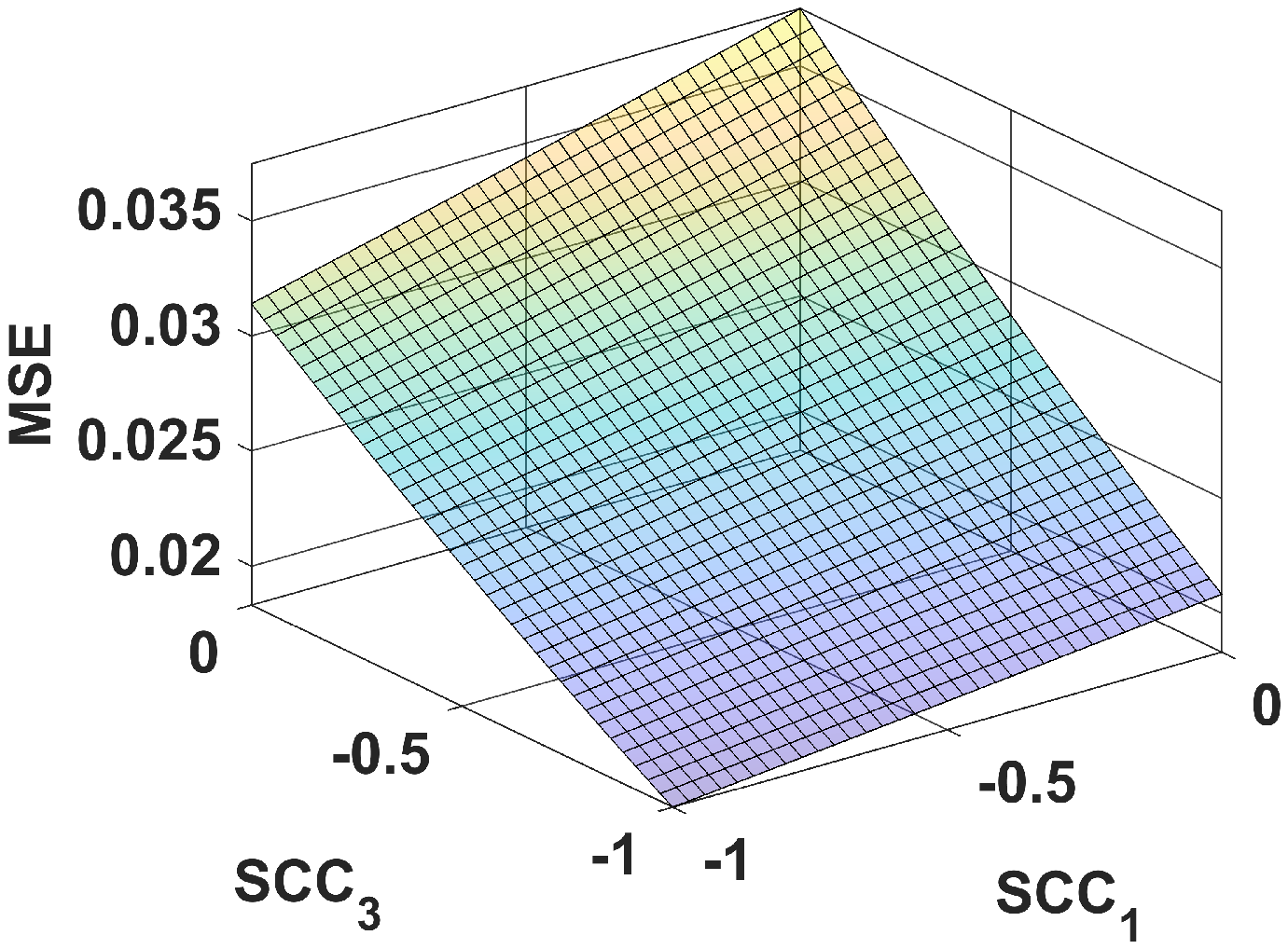}
         \caption{$p_e=0.20$}
         \label{Gate 1 and 4 faulty with p_e=0.2 varying SCC_2 and SCC_3}
     \end{subfigure}
     \begin{subfigure}[b]{0.195\textwidth}
         \centering
         \includegraphics[width=\textwidth]{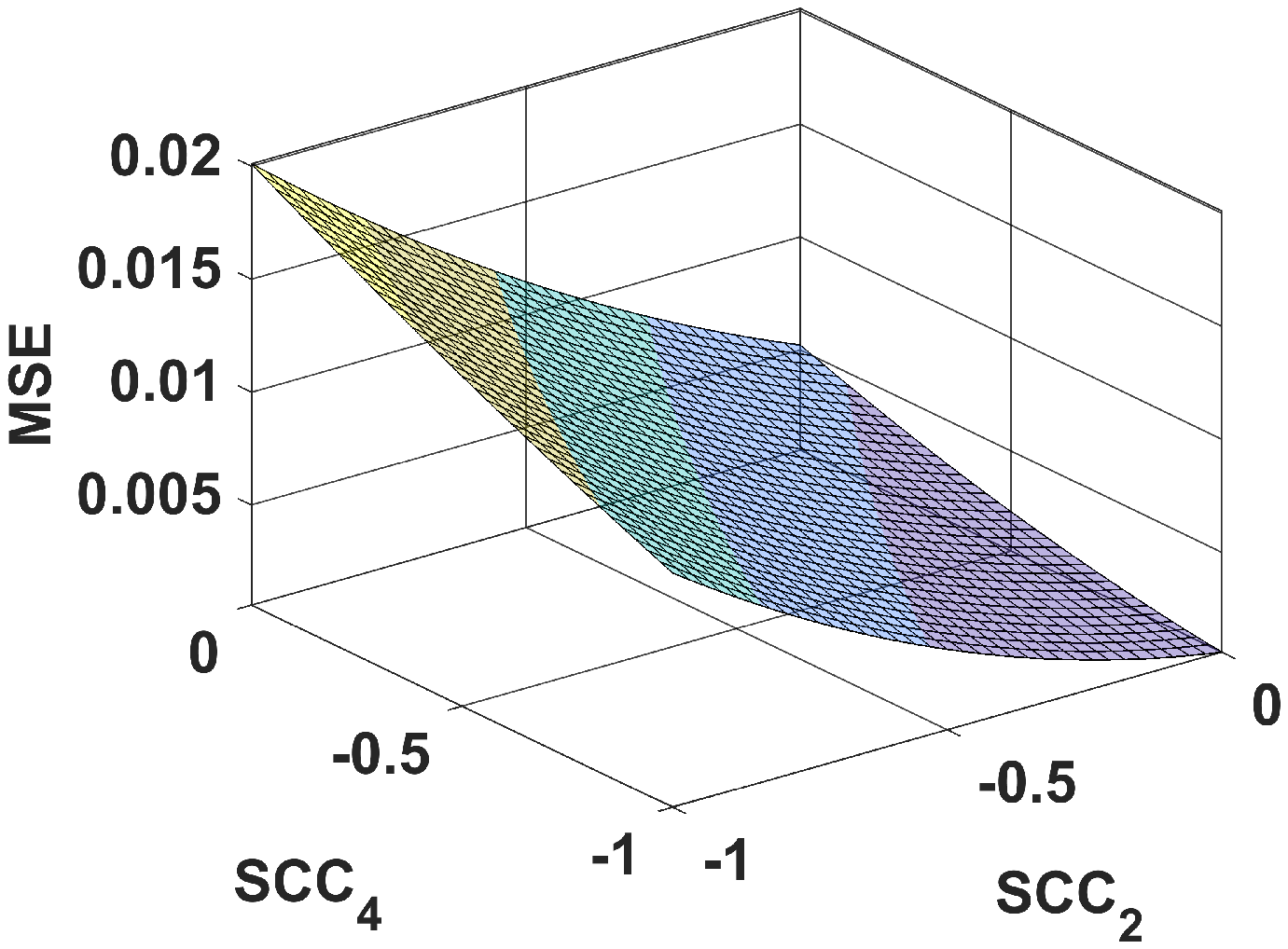}
         \caption{$p_e=0.325$}
         \label{fig:mimo dual scc}
     \end{subfigure}
     \caption{Error minimization using dual ReCo for different error rates;  (a),(b),(c) Gates $4,6$ faulty and (d) Gates $1,4,6$ faulty in Fig. \ref{fig:Intermediate_gate_miso}; (e) Gates $2,4$ faulty in Fig. \ref{fig:Intermediate_gate_MIMO_circuit}. }
     \label{fig:3D_curves_SCC_1_and_SCC_3}
\end{figure*}

 It is interesting to note that for $G2,G4$ faulty with an error rate $\leq 0.2$ the MSE at the output $Z_2 $ can be reduced to $0$  by doing \textit{ReCo} at either of the faulty gates $SCC_2$ or $SCC_4$ as shown in Fig.\ref{fig:Reco_Varying_single_SCC} (c),(d).  We can choose to  \textit{ReCo} at any of the faulty gates without considering the precedence rule. But for  $p_e> 0.2$ the MSE can only be reduced to $0$ with the condition guided by $Algorithm 3$. Modelling $SCC_3$ can reduce MSE to $0$  for error rates up to $0.3$. At $p_e=0.325$, we invoke dual SCC ($SCC_4,SCC_2$) to observe the error resilient behaviour of the circuit as shown in Fig. \ref{fig:3D_curves_SCC_1_and_SCC_3}(e). Thus, \textit{ReCo} analysis only at inputs of every faulty gate is done to avoid the generation of faulty output at a node that is preceded by  non-erroneous outputs. Thus,  modelling of SCCs at the primary gates $G1$ is avoided without perturbing the output logic level at $Z_1$. As the number of faulty gates increase,  the number of correlator circuits also increases. This is contrary to the  selection criteria in $Algorithm 2$. It attempts to  minimize the number of correlator circuits for larger error rates $0.30$ ($l=1$). It is observed that $MSE$ can be reduced to zero for $p_e=0.325$ using dual \textit{ReCo} analysis. The results of analysis are registered in Table II.

 \begin{figure}
 \centering
        \includegraphics[width=0.7\textwidth]{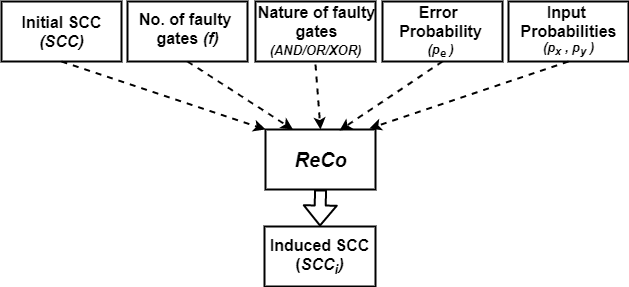}
    \caption{Factors influencing the induced SCC.}
    \label{fig:ReCo_circuit_AND_OR_XOR_gate}
\end{figure}

\begin{table}[]
\caption{Comparison of MSE for \textit{ReCo} at different locations in the circuit in Fig. \ref{fig:Intermediate_gate_miso}. $MSE_{min}$ represents minimum MSE. }
\label{tab:Table1}
\resizebox{0.8\textwidth}{!}{%
\begin{tabular}{|c|c|c|c|c|c|c|c|}
\hline
\multirow{2}{*}{} & \multirow{2}{*}{\begin{tabular}[c]{@{}c@{}}\textbf{Faulty} \\ \textbf{gate(s)}\end{tabular}} & \multirow{2}{*}{\begin{tabular}[c]{@{}c@{}}\textbf{MSE} \\ \textbf{Without    ReCo}\end{tabular}} & \multicolumn{5}{c|}{\textbf{With ReCo (Single/Dual gate)}} \\ \cline{4-8} 
 &  &  &  & $SCC_1$ & $SCC_3$ & $SCC_1$, $SCC_3$ & $SCC_2$, $SCC_3$ \\ \hline
\multirow{8}{*}{\rotatebox{90}{\textbf{Error = 0.125}}} & \multirow{2}{*}{Gate 1} & \multirow{2}{*}{0.00042} & $SCC_i$ & -0.36 & -0.61 & -0.19, -0.16 & -0.26, 0.33 \\ \cline{4-8} 
 &  &  & $MSE_{min}$ & 0 & 0 & 0 & 0 \\ \cline{2-8} 
 & \multirow{2}{*}{Gate 4} & \multirow{2}{*}{0.00085} & $SCC_i$ & -0.51 & -0.45 & -0.27, -0.23 & -0.34, -0.38 \\ \cline{4-8} 
 &  &  & $MSE_{min}$ & 0 & 0 & 0 & 0 \\ \cline{2-8} 
 & \multirow{2}{*}{Gate 1, 4} & \multirow{2}{*}{0.002} & $SCC_i$ & -1 & -0.61 & -0.4471, -0.4001 & -0.51, -0.82 \\ \cline{4-8} 
 &  &  & $MSE_{min}$ & 0 & 0 & 0 & 0 \\ \cline{2-8} 
 & \multirow{2}{*}{\begin{tabular}[c]{@{}c@{}}Gate 1,\\  4, 6\end{tabular}} & \multirow{2}{*}{0.016} & $SCC_i$ & -1 & -1 & -0.9,-0.9 & -1,0 \\ \cline{4-8} 
 &  &  & $MSE_{min}$ & 0.0095 & 0 & 0 & 0.001 \\ \hline
\multirow{8}{*}{\rotatebox{90}{\textbf{Error = 0.20}}} & \multirow{2}{*}{Gate 1} & \multirow{2}{*}{0.0011} & $SCC_i$ & -0.717 & -0.4896 & -0.36,-0.27 & -0.5,-0.49 \\ \cline{4-8} 
 &  &  & $MSE_{min}$ & 0 & 0 & 0 & 0 \\ \cline{2-8} 
 & \multirow{2}{*}{Gate 4} & \multirow{2}{*}{0.0022} & $SCC_i$ & -1 & -0.6385 & -0.45,-0.41 & -1,-0.1 \\ \cline{4-8} 
 &  &  & $MSE_{min}$ & 0 & 0 & 0 & 0 \\ \cline{2-8} 
 & \multirow{2}{*}{Gate 1, 4} & \multirow{2}{*}{0.0044} & $SCC_i$ & -1 & -0.8078 & -0.546, -0.7 & -0.524, -0.68 \\ \cline{4-8} 
 &  &  & $MSE_{min}$ & 0.0015 & 0 & 0 & 0 \\ \cline{2-8} 
 & \multirow{2}{*}{\begin{tabular}[c]{@{}c@{}}Gate 1, \\ 4, 6\end{tabular}} & \multirow{2}{*}{0.0375} & $SCC_i$ & -1 & -1 & -1,-1 & -1, -0.15 \\ \cline{4-8} 
 &  &  & $MSE_{min}$ & 0.0314 & 0.0108 & 0 & 0.01 \\ \hline
\multirow{8}{*}{\rotatebox{90}{\textbf{Error = 0.25}}} & \multirow{2}{*}{Gate 1} & \multirow{2}{*}{0.0017} & $SCC_i$ & -1 & -0.5813 & -0.44, -0.38 & -0.47,-0.83 \\ \cline{4-8} 
 &  &  & $MSE_{min}$ & 0 & 0.0114 & 0 & 0 \\ \cline{2-8} 
 & \multirow{2}{*}{Gate 4} & \multirow{2}{*}{0.0034} & $SCC_i$ & -1 & -0.743 & -52,-0.55 & -1,-0.1 \\ \cline{4-8} 
 &  &  & $MSE_{min}$ & 0.0004 & 0 & 0 & 0 \\ \cline{2-8} 
 & \multirow{2}{*}{Gate 1, 4} & \multirow{2}{*}{0.049} & $SCC_i$ & -1 & -0.8987 & -1,-1 & -1, -0.09 \\ \cline{4-8} 
 &  &  & $MSE_{min}$ & 0.036 & 0 & 0 & 0.0045 \\ \cline{2-8} 
 & \multirow{2}{*}{\begin{tabular}[c]{@{}c@{}}Gate 1, \\ 4,  6\end{tabular}} & \multirow{2}{*}{0.1317} & $SCC_i$ & -1 & -1 & -1,-1 & -1,0 \\ \cline{4-8} 
 &  &  & $MSE_{min}$ & 0.041 & 0.023 & 0.01 & 0.02 \\ \hline
\end{tabular}
}
\end{table}
\begin{table}[]
\centering
\label{tab:MSE_SCC}
\caption {Comparison of MSE for \textit{ReCo} at different locations of circuit in Fig. \ref{fig:Intermediate_gate_MIMO_circuit}}
\resizebox{0.8\textwidth}{!}{%
\begin{tabular}{|c|c|c|c|c|c|c|}
\hline
\multirow{2}{*}{} & \multirow{2}{*}{\textbf{Faulty gates}} & \multirow{2}{*}{\textbf{\begin{tabular}[c]{@{}c@{}}MSE\\ Without  ReCo\end{tabular}}} & \multicolumn{4}{c|}{\textbf{With ReCo (single or dual gate(s))}} \\ \cline{4-7} 
 &  &  & \textbf{} & \textbf{$SCC_2$} & \textbf{$SCC_4$} & \textbf{$SCC_2, SCC_4$} \\ \hline
\multirow{2}{*}{\textbf{Error = 0.25}} & \multirow{2}{*}{\begin{tabular}[c]{@{}c@{}}Gates\\ 2,4\end{tabular}} & \multirow{2}{*}{0.003} & $SCC_i$ & -0.9 & -0.33 & -0.2,-0.4 \\ \cline{4-7} 
 &  &  & $MSE_{min}$ & 0.001 & 0 & 0 \\ \hline
\multirow{2}{*}{\textbf{Error= 0.30}} & \multirow{2}{*}{\begin{tabular}[c]{@{}c@{}}Gates \\ 2,4\end{tabular}} & \multirow{2}{*}{0.004} & $SCC_i$ & -0.93 & -0.87 & -0.15,-1 \\ \cline{4-7} 
 &  &  & $MSE_{min}$ & 0.002 & 0 & 0 \\ \hline
\end{tabular}
}
\end{table}

\section{Case study: Contrast enhancement in  images}
\begin{figure*}{}
    \centering
    \includegraphics[width=\textwidth]{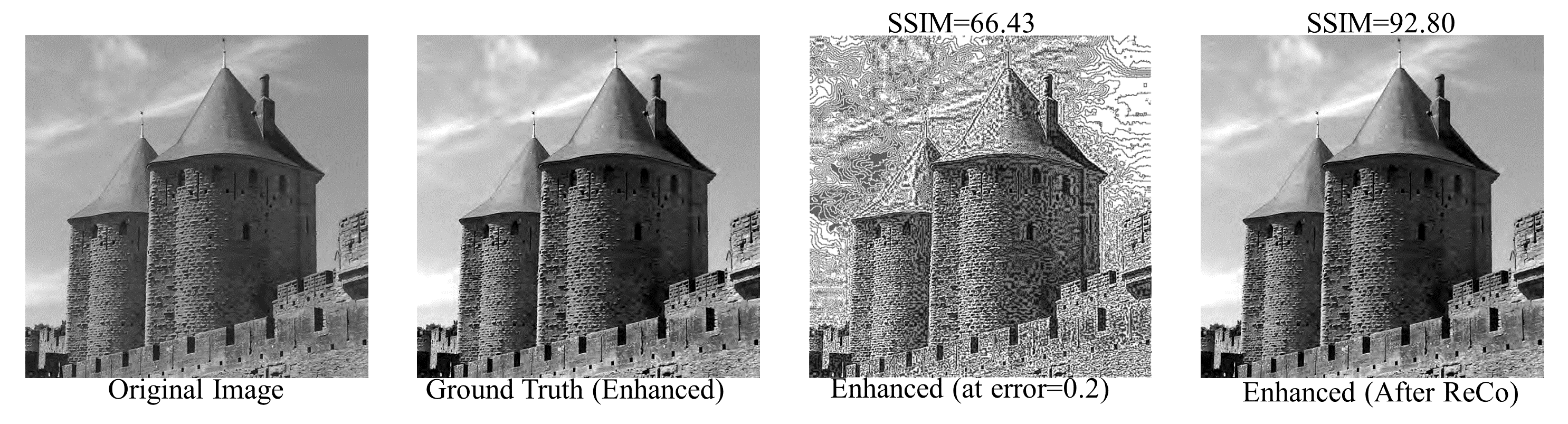}
    \caption{Contrast Stretching operations on the images; \cite{qureshi_contrast_2017} ground truth, with error, and correction using  the proposed methodology.}
    \label{fig:ImageReco}
\end{figure*}

 We have implemented the proposed technique for contrast enhancement \cite{Dhawan1986}  on images to establish the practicality and effectiveness of the proposed scheme. A random image has been taken from the standard dataset \cite{qureshi_contrast_2017}. A transient error at the rate of $0.2$ is given to certain gates to study the effect of the error on the image. Structural Similarity  Index(SSIM)\cite{SSiM1284395} is calculated to measure the similarity between the enhanced images and the ground truth image. It is observed that SSIM of the enhanced image using contrast stretch technique  is considerably low, i.e, $66.43$ when exposed to errors. Exploiting the priority-based approach two SLEs are selected for obtaining an error-resilient behaviour of the circuit.  From Fig. \ref{fig:ImageReco}(d), it is observed that the SSIM index using the proposed \textit{ReCo} method is $92.80$ and is considerably higher compared to the enhanced image with the error which denotes that the proposed  methodology gives a faithful result even in transient error scenarios. Thus the proposed  methodology can be implemented with lower hardware cost in various image processing applications.

\section{Experimental Results and Discussion}

We can now identify the key factors on which the whole analysis is hinged upon. The magnitude and polarity of induced correlation and also the number ($l$) of \textit{ReCo} blocks  depend on several underlying factors. The predominant factor is the amount of transient error in the circuit. As transient error increases the MSE increases exponentially (see graphs in Fig. \ref{fig:MSE_AND_GATE_graphs}(a)). With higher MSE  the value of $l$ tends to be larger. For  $G1,G4,G6$ faulty at $p_e\leq 0.125$ error can be reduced to $0$ using a single \textit{ReCo} block ($SCC_3$). But the same with $p_e \geq 0.20$ error can be reduced to $0$ with $l=2$ as shown in Table \ref{tab:Table1}. This is because MSE is higher in the second case.
 \begin{figure}{}
    \centering
    \includegraphics[width=0.8\textwidth]{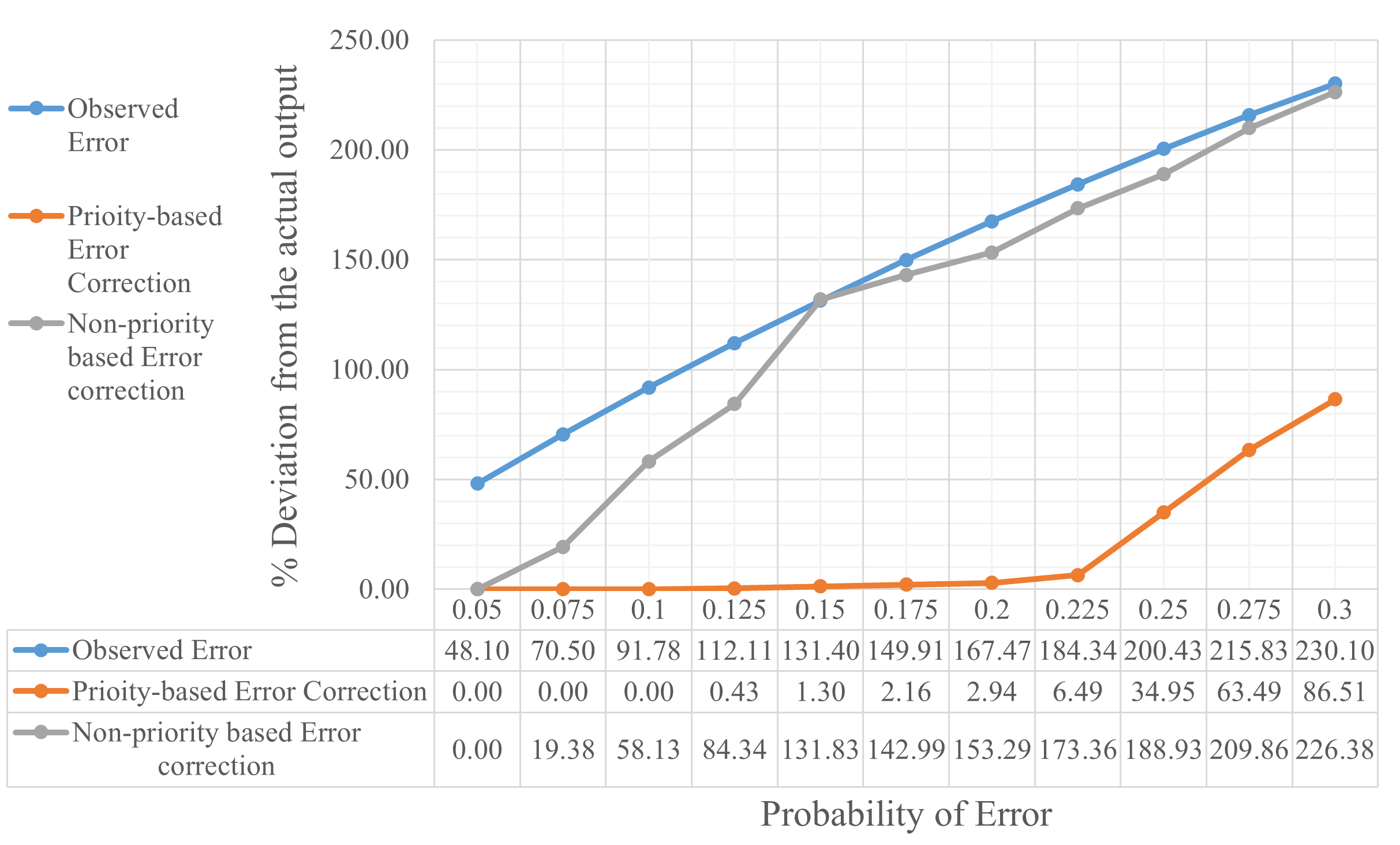}
    \caption{Deviation of two correction approaches from the actual value (without error) of Fig. \ref{fig:Intermediate_gate_miso}.}
    \label{fig:my output_deviation}
\end{figure}
\par It is identified that as the number of faulty gates in the circuit  are increased, the value of $l$ is increased to commensurate with the increased MSE. When  $G1$ faulty and   $G1,G4,G6$ faulty with the same error rate, i.e., $0.20$, the  error-free output is obtained respectively with $l=1$ ($SCC_3/SCC_1$) and $l=2$ ($SCC_3,SCC_1$). Also in the input panel,  the proliferation of the highest priority SLE enhances the  possibility of reducing the error to zero. 
Thus, when  AND gate $G1$ is replaced by an XOR gate, then MSE can be reduced to $0$, even if,  $G1,G4,G6$ are faulty at $p_e=0.25$. 
 \par There is a strong dependence of overall MSE on the nature of faulty gates. If we replace AND gate $G4$ with an XOR gate , the overall MSE is reduced from $0.0375$ to $0.02$, when $G1,G4,G6$ are faulty at $p_e=0.2$. This is because XOR is least sensitive to transient errors. The  position of faulty gates in the circuit has also an impact on the overall MSE. A faulty gate  distant  from the input periphery and closer to the output produces larger MSE. When $G1$  faulty at a rate $p_e=0.2$ gives MSE as $0.0011$, whereas  $G6$ faulty at the same error rate gives larger MSE ($0.0237$). It is also implicit from Fig. \ref{fig:MSE_AND_GATE_graphs}(c),(d) and Fig. \ref{fig:MSE_XOR_GATE_graphs}(c),(d) that the initial assumption of SCC also plays a significant role in determining the exact operating point of SCC for the circuit. In the current experimental setup as in Fig. \ref{fig:Intermediate_gate_MIMO_circuit} the error rates up to $0.3$ can be handled accurately and this is guided by the number of faulty gates in the circuit $(\leq 2) $ and  location of faulty gates (closer to the input side) as shown in Fig. \ref{fig:mimo dual scc}. A block diagram showing the interdependence of these parameters is shown in Fig. \ref{fig:ReCo_circuit_AND_OR_XOR_gate}.
\par  We  have introduced a priority-based selection scheme of SLEs  for larger circuits in Section IV with multiple faulty gates and got some encouraging results. XOR gate exhibits precedency in the correlation-sensitivity list and is considered as a prime element in our analysis. It is inferred that to observe minimum $MSE$ using priority-based approach, minimum number of \textit{ReCo} blocks required is $l=1$. The only exception is \textit{$G1,G4,G6$} faulty at $p_e=0.25$, where two \textit{ReCo} blocks are required to achieve minimum $MSE$. The graphs of Fig. \ref{fig:my output_deviation} are obtained with $G1,G4,G6$ faulty at different error rates. The deviation in output from the actual value (without error) using the proposed scheme is much less compared to the non-priority based approach.  It is observed that this approach is able to handle high error rates with reduced number of correlator circuits. The number of iterations required to achieve the desired value is also less compared to the non-priority based approach. The priority-based (red) deviation graph is obtained with one \textit{ReCo} block whereas the  deviation with non-priority based approach (gray) is obtained with two \textit{ReCo} blocks to model the output. The blue line correspond to the observed error (without \textit{ReCo}). This  shows  the efficacy of the proposed priority-based approach in terms of hardware design.
\section{CONCLUSION}
Recent applications of stochastic computing have involved noisy operating conditions leading to incorrect results at times. The source of inaccuracy has been predominantly traced to transient errors. In this paper, we have progressively varied the transient error probabilities for single gates and observed its effect on the MSE of these gates. Attempts are made to formulate the process within a mathematical framework. In view of the effect of varying correlation on the MSE, we advanced our study into realistic multi-level circuits where single or multiple gates may be prone to transient errors. For such circuits, we have developed the \textit{ReCo} framework to minimize the overall MSE. Algorithm $2$ introduces a priority-based approach of choosing SLE to reduce the number of correlator circuits and to obtain the desired level of accuracy quickly under noisy operating conditions. Inevitably, there are conflicts in constraints in different applications which is handled using different approaches. Algorithm $3$ recognizes and eliminates this ambiguity for a MIMO circuit by introducing correction blocks at fault specified nodes only. Both these algorithms are observed to handle errors and yield accurate results even at high transient error rates. In our future work, we will explore other variants of these algorithms to achieve better correction using lesser \textit{ReCo} blocks. Also, we will try to develop some equivalent form of  combinational and sequential circuits with a given set of conditions such that the task of error minimization is achieved at a lesser hardware cost and low latency. 

\begin{algorithm}[h]
\begin{algorithmic}[1]
\caption{\textit{ReCo analysis for MISO Circuits}} 
\label{Algorithm:Miso}
\STATE \textbf{Input:} $CIRCUIT$ \textit{{// Circuit with n number of inputs.}}
\\\textbf{Output: }$MSE, SCC_i $
\STATE \textbf{Variable Initialization:} 
\\$p\_arr[\;]=\{XOR,OR,AND\}$ \textit{//Priority Sequence} \\
$F\_gate[\;]=\{0\}\;\;$//{ number of faulty gates}\\
$I\_gate[\;]=\{ number\; of\; input\; gates\}$ \\
$FC\_I\_gate[\;]=\{0\}$ \textit{//  Input gates $\in F\_gate[\;]$}\\
$S\_FC\_I\_gate[\;]=\{0\}$ \textit{// Sorted $FC\_I\_gate[\;]$}
\STATE {$F\_gate$=\textit{\textbf{FaultEvaluation}}($CIRCUIT$)}\\ \hspace{12pt} \textit{// Identify faulty gates in the circuit }
\STATE {$FC\_I\_gate$=\textit{\textbf{IsConnected}}($I\_gate, F\_gate)$}
\STATE{$S\_FC\_I\_gate$=\textit{\textbf{PrioritySort}}($FC\_I\_gate$, $p\_arrr)$ 

\STATE
\For{$j=1 \: \text{to} \: j<=\textit{\textbf{maxElement}}$ ($S\_FC\_I\_gate)$ }
{
$[L\_MSE[j],SCC_i[j]]= \textit{\textbf{ReCo}}(S\_FC\_I\_gate[j]$)\\ 
\If{($L\_MSE[j] <= \delta$)}{ return $L\_MSE[j], SCC_i[j]$}
}
\STATE $[S\_L\-gate, L\_MSE]=\textbf{\textit{Sort}}(L\_MSE)$ //\textit{Sort MSE value along with their gate number}
\STATE \For{$j=1 \: \text{to} \: j<= \textit{\textbf{maxElement}}($S\_FC\_I\_gate$)$}
{[$MSE[j], SCC_i[j]]=NewReco( S\_L\_gate,j+1)$\\
\If{($MSE[j] <= \delta$)}{ return $MSE[j], SCC_i[j]$}

}}
\STATE { return $\argmin_{MSE[j]}$  \{$MSE[j]$ ,  $SCC_i[j]$\}}
\end{algorithmic}
\end{algorithm}

\begin{algorithm}[h]
\begin{algorithmic}[1]
\caption{\textit{ReCo analysis for MIMO Circuits}} 
\label{Algorithm:Miso}
\STATE \textbf{Input:} $CIRCUIT$ \textit{{// Circuit with n number of inputs}}
\\\textbf{Output: }$MSE, SCC_i $
\STATE \textbf{Variable Initialization:} 
\\$p\_arr[\;]=\{XOR,OR,AND\}$ \textit{//Priority Sequence} \\
$F\_gate[\;]=\{0\}\;\;$//{ number of faulty gates}\\
$S\_F\_gate[\;]=\{0\}$ \textit{// Sorted $F\_gate[\;]$}
\STATE {$F\_gate$=\textit{\textbf{FaultEvaluation}}($CIRCUIT$)}\\ \hspace{12pt} \textit{// Identify faulty gates in the circuit }
\STATE{$S\_F\_gate$=\textit{\textbf{PrioritySort}}($F\_gate$, $p\_arrr)$ 

\STATE
\For{$j=1 \: \text{to} \: j<= \textit{\textbf{maxElement}}($S\_F\_gate$)$}
{
$[L\_MSE[j],SCC_i[j]]= \textit{\textbf{ReCo}}(S\_F\_gate[j]$)\\ 
\If{($L\_MSE[j] <= \delta$)}{ return $L\_MSE[j], SCC_i[j]$}
}
\STATE $[S\_L\-gate, L\_MSE]=\textbf{\textit{Sort}}(L\_MSE)$ //\textit{Sort MSE value along with their gate number}
\STATE \For{$j=1 \: \text{to} \: j<= \textit{\textbf{maxElement}}($S\_F\_gate$)$}
{[$MSE[j], SCC_i[j]]=NewReco( S\_L\_gate,j+1)$\\
\If{($MSE[j] <= \delta$)}{ return $MSE[j], SCC_i[j]$}


}}
\STATE { return $\argmin_{MSE[j]}$  \{$MSE[j]$ ,  $SCC_i[j]$\}}
\end{algorithmic}
\end{algorithm}

\bibliographystyle{unsrt}
\bibliography{bibliography01.bib}

\vspace{12pt}
\color{red}

\end{document}